\documentclass[10pt]{article}

\usepackage{graphicx}              
\usepackage{amsmath}               
\usepackage{amssymb, amsbsy, latexsym} 
\usepackage{amsfonts}              
\usepackage[mathscr]{eucal}     
\usepackage{amsthm}                
\usepackage{multirow}               
\usepackage{txfonts}

\DeclareMathSymbol{\C}{\mathbin}{AMSb}{"43}
\DeclareSymbolFont{AMSb}{U}{msb}{m}{n}

\makeatletter
\@addtoreset{equation}{section}
\makeatother

\newtheorem{Theorem}{Theorem}[section]
\newtheorem{Definition}{Definition}[section]

\newtheorem{Corollary}{Corollary}[section]

\def\be{\begin{eqnarray}}
\def\ee{\end{eqnarray}}

\newcommand{\cd}{\mathcal D}
\newcommand{\ce}{\mathcal E}
\newcommand{\cf}{\mathcal F}

\newcommand{\ch}{\mathcal H}
\newcommand{\ci}{\mathcal I}
\newcommand{\cj}{\mathcal J}

\newcommand{\cl}{\mathcal L}
\newcommand{\cm}{\mathcal M}

\newcommand{\ct}{\mathcal T}

\newcommand{\cv}{\mathcal V}

\newcommand{\cz}{\mathcal Z}

  \newcommand{\Fa}{\mathfrak{A}}

  \newcommand{\Fd}{\mathfrak{D}}

  \newcommand{\Fn}{\mathfrak{N}}

\renewcommand{\a}{\alpha}
\renewcommand{\b}{\beta}
\newcommand{\g}{\gamma}
\newcommand{\G}{\Gamma}

\newcommand{\eps}{\epsilon}

\newcommand{\Sig}{\Sigma}
\renewcommand{\l}{\lambda}
\renewcommand{\L }{\Lambda}
\renewcommand{\o}{\omega}
\renewcommand{\O}{\Omega}
\renewcommand{\t}{\tau}

\newcommand{\rmd}{\mathrm d}

\newcommand{\peq}{^\Psi\!\!\!\!\!=}
\newcommand{\lag}{\langle}
\newcommand{\rag}{\rangle}
\newcommand{\lt}{\left}
\newcommand{\rt}{\right}

\begin{document}

\title{\sf Canonical Path-Integral Measures for Holst and Plebanski Gravity:\\
II. Gauge Invariance and Physical Inner Product}

\author{
{\sf Muxin Han}\thanks{{\sf 
mhan@aei.mpg.de}}\\
\\
{\sf MPI f. Gravitationsphysik, Albert-Einstein-Institut,} \\
           {\sf Am M\"uhlenberg 1, 14476 Potsdam, Germany}\\
\\
{\sf Institut f. Theoretische Physik III, Universit\"{a}t Erlangen-N\"{u}rnberg} \\
{\sf Staudtstra{\ss}e 7, 91058 Erlangen, Germany}
}
\date{}
\maketitle

\thispagestyle{empty}

\begin{abstract} 
{\sf This article serves as a continuation for the discussion in \cite{EHT}, we analyze the invariance properties of the gravity path-integral measure derived from canonical framework, and discuss which path-integral formula may be employed in the concrete computation e.g. constructing a spin-foam model, so that the final model can be interpreted as a physical inner product in the canonical theory.

The present article is divided into two parts, the first part is concerning the gauge invariance of the canonical path-integral measure for gravity from the reduced phase space quantization. We show that the path-integral measure is invariant under all the gauge transformations generated by all the constraints. These gauge transformations are the local symmetries of the gravity action, which is implemented without anomaly at the quantum level by the invariant path-integral measure. However, these gauge transformations coincide with the spacetime diffeomorphisms only when the equations of motion is imposed. But the path-integral measure is not invariant under spacetime diffeomorphisms, i.e. the local symmetry of spacetime diffeomorphisms become anomalous in the reduced phase space path-integral quantization.

In the second part, we present a path-integral formula, which formally solves all the quantum constraint equations of gravity, and further results in a rigging map in the sense of refined algebraic quantization (RAQ). Then we give a formal path-integral expression of the physical inner product in loop quantum gravity (LQG). This path-integral expression is simpler than the one from reduced phase space quantization, since all the gauge fixing conditions are removed except the time-gauge. The resulting path-integral measure is different from the product Lebesgue measure up to a local measure factor containing both the spacetime volume element and the spatial volume element. This formal path-integral expression of the physical inner product can be a starting point for constructing a spin-foam model.}

\end{abstract}

\newpage 

\tableofcontents

\newpage

\section{Introduction}

The path-integral measure plays an important role in the path-integral quantization of a dynamical system. However at first look, the choice of path-integral measure is ambiguous. Even for a simple system like a free particle on $\mathbb{R}^3$, all the different measures on the space of paths can play the role as a path-integral measure, and lead to different quantum theories. Usually the way to select a right measure for path-integral is to establish the equivalence with the canonical quantization of the same system. For the simple systems, e.g. a free particle on $\mathbb{R}^3$ and the free fields on the Minkowski spacetime, the correct path-integral measures are simply the formal Lebesgue measures. But for general systems more complicated than these, in particular a constrained system, the path-integral measure is unlikely to be always a Lebesgue meausure. It is suggested by the literatures e.g. \cite{QuantizationGauge} that for a general constrained system, one should use a measure derived from the Liouville measure on the reduced phase space. The reason is that the reduced phase space Liouville measure has direct relation with the canonical quantization in the reduced phase space of the constrained system. The path-integral formulation is of interest in the quantization of general relativity (GR), a theory where space-time covariance plays a key role. The spacetime diffeomorphism of GR results in that GR formulate gravity as a dynamical system with a number of first-class constraints with some certain complications. In quantizing such a complicated dynamical system as GR, one has to be careful about the choice of path-integral measure. 

Currently loop quantum gravity (LQG) is a mathematically rigorous quantization of general relativity that preserves background independence --- for reviews, see \cite{book,rev}. The spin-foam model can be thought of as a path-integral framework for loop quantum gravity, directly motivated from the ideas of path-integral adapted to reparametrization-invariant theories \cite{sfrevs,spinfoam2}. So far, however only the kinematical structure of LQG is used in motivating the spin-foam framework, while the canonical formulation of the dynamics in LQG doesn't contribute to the current spin-foam models. Instead the resulting framework is remarkably close to the path-integral quantization of so-called BF theory, a topological field theory whose quantization is exactly known \cite{bf}. GR can in fact be formulated as a contrained BF theory, yielding the Plebanski formulation \cite{plebanski}, and for this reason the Plebanski formulation is usually the starting point for deriving the dynamics of spin-foams. However, the consistency between spin-foam model and canonical framework of LQG hasn't been well understood. And the path-integral measure consistent with canonical theory hasn't been incorporated in the current spin-foam models yet.  

In \cite{EHT}, we derive two path-integral formulations for both the Holst and Plebanski-Holst actions from the reduced phase space quantization. Thus we show that their path-integral measures are consistent with the canonical theory, so they are candidates for the spin-foam construction. However, there are immediately two questions concerning the resulting path-integral formulae:

\begin{enumerate}

\item The first question is a conceptual question: The resulting path-integral measure\footnote{In a path-integral formula $\int\cd\mu\ e^{iS}$, some author use the term ``a path-integral measure'' referring to $\cd\mu$, but some others use it referring to $\cd\mu\ e^{iS}$. We are following the first convention. But there is no different between these two convention when we are consider the invariance of the path-integral measure under the symmetries of the action $S$.} is not a formal Lebesgue measure, but with a so called, local measure factor of the shape $\cv^nV_s^m$, where $\cv$ is the spacetime volume element and $V_s$ is the spatial volume element. And the powers $m,n$ are different between the cases of the Holst action and the Plebanski-Holst action. The appearance of spatial volume element breaks the manifest spacetime diffeomorphism invariance of the path-integral measure, which leads to the first question: What is the implication of this diffeomorphism non-invariant path-integral measure? Does it mean that this path-integral quantization of gravity breaks the spacetime diffeomorphism invariance? 

\item The second one is a practical question: In the path-integral formulae in \cite{EHT} given by reduced phase space quantization, the integrands contain several gauge fixing conditions, one for each first-class constraints. This fact reflects that we are considering the quantization of a gauge system. For the conventional computation of the path-integral amplitude, one often need to introduce the ghost fields and write down an effective action. However, if we consider the background independent quantization for GR such as the spin-foam models, the gauge fixing terms are too complicated to be implemented. Then the question is: Can we find some ways to circumvent the gauge fixing conditions, in order to make path-integrals in \cite{EHT} computable? 

\end{enumerate}
 
Our answer for the first question is: If our path-integral quantization is consistent with the canonical theory of gravity, then the local symmetry of spacetime diffeomorphism is broken in the quantum level. The reason is the following: It turns out in \cite{wald} that the local symmetry group Diff(M) corresponding to the spacetime diffeomorphisms is not projectable under the projection map from the space of metric to the phase space. There doesn't exist any canonical generator on the phase space generating spacetime diffeomorphisms for the phase space variables. Thus Diff(M) is not the group of gauge symmetries in the canonical GR. A direct consequence is that the first-class constraints (the spatial diffeomorphism constraint and the Hamiltonian constraint) generate a constraint algebra which is not a Lie algebra, i.e. the structure functions appears. Therefore the gauge transformations doesn't form a group in the canonical theory of gravity, since they are generated by the first-class constraints. The collection of the gauge transformations is at most an enveloping algebra, whose generic element is a product of infinitesimal gauge transformations. We refer to this enveloping algebra as the ``Bergmann-Komar group'' BK(M) \cite{BK}. It coincides with Diff(M) only when the equation of motion is imposed. This Bergmann-Komar group essentially determines the dynamical symmetry of canonical GR, while Diff(M) is only the kinematical symmetry of the theory. Here by kinematical we mean that the symmetry group is insensitive to the form of the Lagrangian. This point can be illustrated by comparing the Einstein-Hilbert action with the high-derivative action
\be
\int\rmd^4x\ \sqrt{\big|\det g(x)\big|}\ R_{\a\b\g\delta}(x)\ R^{\a\b\g\delta}(x).
\ee
Both the actions are spacetime diffeomorphism invariant, but their dynamics are dramatically different, which can be seen from their constraints. 

Therefore we can see that a path-integral quantization of GR with the local symmetries of Diff(M) cannot be consistent with the canonical theory with the gauge symmetries of BK(M). They are consistent at most in the semiclassical limit. Then an immediate question is whether the dynamical symmetry of the Bergmann-Komar group BK(M) is implemented in our path-integral quantization in \cite{EHT}. The answer is positive. We will show in the present paper that the Bergmann-Komar group BK(M), which is also a collection of local symmetry of the gravity action, is implemented anomaly-freely in the path-integral quantization derived from reduced phase space quantization. Therefore all the informations of the dynamical gauge symmetry have been incorporated in this path-integral quantization. Moreover, if we approximate the path-integral around a classical solution of the equation of motion, the semiclassical limit recovers the spacetime diffeomorphism invariance.

Since our path-integral measure is derived from canonical framework of GR, instead of asking it to be diffeomorphism invariant, one should rather ask whether it can solve all the quantum constraint equations. More precisely, given the kinematical Hilbert space $\ch_{Kin}$ of GR (which can be realized by the kinematical Hilbert space in LQG), we represent the classical constraints $C_I$ to be operators $\hat{C}_I$ on the kinematical Hilbert space. Then the quantum constraint equations are $\hat{C}_I\Psi=0$. A correct path-integral formula should gives a rigging map for the refined algebraic quantization (RAQ) \cite{RAQ}, mapping the kinematical states in a dense domain of $\ch_{Kin}$ to the space of solutions of the quantum constraint equations. In the present paper, we will show that the path-integral formula derived from reduced phase space quantization does give the desired rigging map (will be denoted by $\eta_\o$), which formally solves all the (Abelianized) constraints of GR quantum mechanically. Finally we can write down formally a physical inner product of LQG in terms of this path-integral formula. This result means that this path-integral formula correctly represents the quantum dynamics of GR.

Our resulting path-integral expression of the physical inner product also effectively answer the second question above. It turns out that all the gauge fixing conditions are removed in the path-integral representing the physical inner product, except the so called, time-gauge. For example, we will show that the physical inner product can be formally represented by a path-integral of the Plebanski-Holst action
\be
\lag\eta_\o(f')|\eta_\o(f)\rag_{Phys}&=&\frac{Z_T(f,f')}{Z_T(\o,\o)}\nonumber\\
Z_T(f,f')&=&\int_{II\pm}\cd\o_\a^{IJ}\cd B_{\a\b}^{IJ} \prod_{x\in M} \cv^{13/2} V^{9}_s\ \delta^{20}\left(\eps_{IJKL}\ B_{\a\b}^{IJ}\ B_{\g\delta}^{KL}-\frac{1}{4!}\cv\eps_{\a\b\g\delta}\right)\ \delta^3(T_c)\nonumber\\
&&\times\lt[\exp i\int_M B^{IJ}\wedge (F-\frac{1}{\g}* F)\rt]\ \overline{f\lt(A_a^i\rt)_{t_f}}f'\lt(A_a^i\rt)_{t_i}\label{1}
\ee
where $\cv$ and $V_s$ are spacetime volume element and spatial volume element respectively. The time-gauge $T_c$ has to be there since our analysis starts from the Ashtekar-Barbero-Immirzi Hamiltonian formulation \cite{ABI}, which is the starting point of LQG. Since the gauge fixing conditions disappear, this physical inner product represented by the path-integral is ready for the concrete computation, by employing the technique of the spin-form model. The resulting spin-foam model will possess a direct canonical interpretation as a physical inner product.

One can see from Eq.(\ref{1}) that, as we expected, the local measure factor $\cv^{13/2} V^{9}_s$ appears in the physical inner product, it actually reflects the fact that we are considering a dynamical system with constraints. The quantum effect of this type of local measure factor has been discussed in the literature since 1960s (see for instance \cite{FV,GT}) in the formalism of geometrodynamics and its {\it background-dependent} quantizations (stationary phase approximation). The outcome from the earlier investigations appears that in the background-dependent quantization, this local measure factor only contributes to a divergent part of the loop-order amplitude, thus their meanings essentially depend on the regularization scheme. Then it turns out that one can always choose certain regularization schemes such that, either the local measure factor never contributes to the transition amplitude (e.g. dimensional regularization), or it is canceled by the divergence from the action \cite{FV,GT}. Thus in the end, the effect from the local measure factor may be ignored in the practical computation of background-dependent quantization.

In the formalism of connection-dynamics for GR, however, when we perform {\it background-independent} quantization like the spin-foam models, in principle the local measure factor should not be simply ignored, because the regularization arguments in background-dependent quantization are not motivated in the background-independent context anymore. For example, the spin-foam models are defined on a triangulation of the spacetime manifold with a finite number of vertices, where at each vertex the value of the local measure factor is finite, and the action also doesn't show any divergence. Thus in principle one has to consider the quantum effect implied by this local measure factor in the context of spin-foam model, if one wants to relate the spin-foam model to the canonical theory. An immediate consequence is that the crossing symmetry in spin-foam model may be broken by the spatial volume measure factor, which is consistent with the canonical LQG in terms of Hamiltonian constraint operator and master constraint operator \cite{QSD}.

The present article is organized as follows: 

In section \ref{generalinv}, we first review the reduced phase space quantization for a general constrained system, and perform the derivation for its path-integral formulation. And give a general argument about in which circumstance the path-integral measure is invariant under the infinitesimal gauge transformations generated by the first-class constraints. 

In section \ref{GRinv}, the general consideration is applied to the case of GR. We analyze the path-integral measures for both the ADM formalism and the Holst action. We show that the path-integral measures is invariant under the Bergmann-Komar group, which is also a collection of the local symmetries. However, the spacetime diffeomorphism symmetries become anomalous in this path-integral quantization. 

In section \ref{PIRAQ}, we first briefly review the general programme of refined algebraic quantization. Then for a general time-reparametrization invariant constrained system, we give a general formal expression of the rigging map by using the path-integral from its reduced phase space quantization. After that we apply the general expression to the case of gravity coupling with 4 real massless scalar field. In the end, we obtain the physical inner product of GR formally represented by a path-integral formula of the Holst action or the Plebanski-Holst action.

\section{A general dynamical system with both first-class and second-class constraints}\label{generalinv}

\subsection{Reduce phase space quantizations of the constraint system}\label{reduce}

We first consider a general reparametrization-invariant dynamical system, whose Hamiltonian is a linear combination of constraints. We will employ the relational framework to construct Dirac observables of the system \cite{observable} and then perform the canonical quantization in reduced phase space. The advantages of this approach are that: (1) We will obtain a direct interpretation for the path-integral amplitude with boundary \emph{kinematical state} as the physical inner poduct between \emph{physical states} in the canonical reduced phase space quantization in terms of relational framework; (2) This approach will also help us in considering the rigging map in refined algebraic quantization, which will be proposed in the section \ref{PIRAQ}.  

We first briefly recall the relational framework of constructing Dirac observables for general covariant systems (see \cite{observable} for details, see also \cite{tina}). First of all, we consider the phase space for a dynamical system $(\cm,\o)$ with a collection of the first-class constraints $C_I$, $I\in\ci$ which is an arbitrary index set. These first-class constraints in the most general case close under the Poisson bracket from $\o$:
\be
\{C_I,C_J\}=f_{IJ}^{\ \ K}C_K
\ee
where $f_{IJ}^{K}$ in general is a structure function. If the dynamical system also has a number of second-class constraints $\Phi_i$, we denote by $\cm$ the constraint surface where all the second-class constraints vanish, and $\o$ is the Dirac symplectic structure on $\cm$. We choose a collection of the gauge variant functions $T^I$ (clock functions), $I\in\ci$, providing a local coordinatization of the gauge orbit $[m]$ of any point $m$ in the phase space, at least in a neighborhood of the constraint surface $\overline{\cm}:=\{m\in\cm\ |\ C_I(m)=0,\ \forall I\in\ci\}$. It follows that the matrix $A_{I}^J:=\{C_I,T^J\}$ must be locally invertible. Consider the equivalent set of constraints 
\be
C'_I:=\sum_J [A^{-1}]^J_IC_J
\ee
the new set of constraints has the properties that $\{C'_I,T^J\}\approx\delta^J_I$ and their Hamiltonian vector fields $X_I:=\chi_{C'_I}$ are weakly commuting. For any real numbers $\b^I$ 
\be
X_\b:=\sum_I\b^I\ X_I.
\ee
Then the gauge transformation can be expressed in terms of formal Taylor series for any function $f$ on the phase space
\be
\a_\b(f):=\exp(X_\b)\cdot f=\sum_{n=0}^\infty\frac{1}{n!}X_\b^n\cdot f
\ee
by the mutually weak commutativity of $X_I$. The idea of relational framework is that by using the gauge variant clock variables $T^I$, we construct a (at least weakly) gauge invariant Dirac observable from each gauge variant phase space function. The resulting Dirac oberservables can separate the points on the reduce phase space. It allows one to coordinatize the reduced phase space by the Dirac observables so constructed. Given a collection of the phase space constants $\t^I$, the weakly gauge invariant Dirac observable associated with the partial observables $f$ and $T^I$ are defined by \cite{observable}
\be
O_f(\t):=[\a_\b(f)]_{\a_\b(T^I)=\t^I}.\label{Of}
\ee
One can check that $\a_\b(T^I)\approx T^I+\b^I$. Notice that after equating $\b^I$ with $\t^I-T^I$, the previously phase space independent quantities $\b$ become phase space dependent, therefore it is important in Eq.(\ref{Of}) to first compute the action of $X_\b$ with $\b^I$ treated as phase space independent and only then to set it equal to $\t^I-T^I$. Therefore on the constraint surface $O_f(\t)$ can be expressed as formal series
\be
O_f(\t)\approx[\a_\b(f)]_{\b^I=\t^I-T^I}=\sum_{\{k_I\}=0}^\infty\prod_{I}\frac{(\t^I-T^I)^{k_I}}{k_I!}\prod_I(X_I)^{k_I}\cdot f
\ee
A significant consequence of the above construction is that the map $O(\t): f\to O_f(\t)$ is a weak Poisson homomorphism (homomorphism only on $\overline{\cm}$) from the Poisson algebra of the functions on the constraint surface defined by $C_I$ ,$T^J$ with respect to the Dirac bracket $\{,\}_D$ to the Poisson algebra of the weak Dirac observables $O_f(\t)$ \cite{observable}. To write the relation explicitly, 
\be
&&O_f(\t)+O_{f'}(\t)=O_{f+f'}(\t),\ \ \ \ \ \ \ \ \ \ \ \ O_f(\t)O_{f'}(\t)\approx O_{ff'}(\t),\nonumber\\
&&\{O_f(\t),O_{f'}(\t)\}\approx\{O_f(\t),O_{f'}(\t)\}_D\approx O_{\{f,f'\}_D}(\t)
\ee 
where the Dirac bracket is explicitly given by 
\be
\{f,f'\}_D=\{f,f'\}-\{f,C_I\}[A^{-1}]^I_J\{T^J,f'\}+\{f',C_I\}[A^{-1}]^I_J\{T^J,f\}.
\ee

Suppose that we can choose the canonical coordinates, such that the clock functions $T^I$ are some of the canonical coordinates, we write down the complete canonical pairs $(q^a,p_a)$ and $(T^I,P_I)$ where $P_I$ is the conjugate momentum of $T^I$. And at least locally one can write the constraints $C_I$ in an equivalent form:
\be
\tilde{C}_I=P_I+h_I(q^a,p_a,T^J)\label{solve}
\ee
thus one can solve the constraints by setting $P_I=-h_I(q^a,p_a,T^J)$. On the constraint surface, the Dirac observable associated with the clocks $T^I$ 
\be
O_{T^I}(\t):=[\a_\b(T^I)]_{\a_\b(T^I)=\t^I}=\t^I
\ee
is simply a constant on the phase space. We also define the Dirac observables associated with the canonical pairs $(q^a,p_a)$
\be
Q^a(\t):=O_{q^a}(\t)\ \ \ \ \ \ \ \ \ \ \ \ P_a(\t):=O_{p_a}(\t)
\ee
and the ``equal-time" Poisson bracket:
\be
\{P_a(\t),Q^b(\t)\}\approx\delta^b_a\ \ \ \ \ \ \ \ \ \ \ \{P_a(\t),P_b(\t)\}\approx\{Q^a(\t),Q^b(\t)\}\approx0
\ee
Thus we can see that for each $\t$ the pairs $(Q^a(\t),P_a(\t))$ form the canonical coordinates of the reduced phase space.

On the other hand, the collection of the constraints $\tilde{C}_I=P_I+h_I(q^a,p_a,T^J)$ forms a strongly Abelean constraint algebra. The reason is the following: $\tilde{C}_I$'s are first class, i.e. $\{\tilde{C}_I, \tilde{C}_J\}=\tilde{f}_{IJ}^{\ \ K}\tilde{C}_K$ for some new structure function $\tilde{f}$. the left hand side is independent of $P_I$, so must be the right hand side. Then the right hand side can be evaluated at any value of $P_I$. So we set $P_I=-h_I$. If we write $C'_I=\sum_JK_{IJ}\tilde{C}_J$ for a regular matrix $K$ and since $\{C'_I,T^J\}\approx\delta^J_I=\{\tilde{C}_I,T^J\}$, we obtain that $K_{IJ}\approx\delta^J_I$ and $C'_I=\tilde{C}_I+O(C^2)$, which means that $C'_I$ and $\tilde{C}_I$ is different by the terms quadratic in the constraints. It follows that the Hamiltonian vector fields $X_I$ and $\tilde{X}_I$ of $C'_I$ and $\tilde{C}_I$ are weakly commuting. We now set $H_I(\t)=H_I(Q^a(\t),P_a(\t),\t):=O_{h_I}(\t)\approx h_I(Q^a(\t),P_a(\t),\t)$. For any function $f$ depending only on $p^a$ and $q_a$, we have the ``equal-time"commutator:
\be
\{H_I(\t), O_f(\t)\}&\approx& O_{\{h_I,f\}_D}(\t)\ =\ O_{\{h_I,f\}}(\t)\ =\ O_{\{\tilde{C}_I,f\}}(\t)
\ =\ \sum_{\{k\}}\prod_{J}\frac{(\t^J-T^J)^{k_J}}{k_J!}\prod_JX_J^{k_J}\cdot \tilde{X}_I\cdot f\nonumber\\
&\approx&\sum_{\{k\}}\prod_{J}\frac{(\t^J-T^J)^{k_J}}{k_J!}\ \tilde{X}_I\cdot\prod_JX_J^{k_J}\cdot f\ \approx \sum_{\{k\}}\prod_{J}\frac{(\t^J-T^J)^{k_J}}{k_J!}\ {X}_I\cdot\prod_JX_J^{k_J}\cdot f\nonumber\\
&=&\frac{\partial}{\partial\t^I}O_f(\t)\label{Hamiltonian}
\ee
which means that the Dirac observable $H_I(\t)$ is a ``time-dependent" generator for the gauge flow on the constraint surface. Thus we call $H_I(\t)$ the (time-dependent) physical Hamiltonian if we are dealing with a general reparametrization-invariant system with vanishing Hamiltonian. Moreover, the algebra of the physical Hamiltonians is weakly Abelean, because the flows $\a^\t: O_f(\t_0)\mapsto O_f(\t+\t_0)$ forms a Abelean group of weak automorphisms.

Then we come to the quantization on the reduced phase space, where all the classical constraint is solved and all the classical elementary observables are invariant under gauge transformation. We start our quantization in Heisenberg picture. On the kinematical level, the quantum algebra $\Fa$ is generated by the gauge invariant observables $\hat{Q}^a(\t)$ and $\hat{P}_a(\t)$ with the ``equal-time" canonical commutation relation for any $\t$ (in particular $\t=0$)
\be
&&[\hat{P}_a(\t),\hat{Q}^b(\t)]=-i\{P_a(\t),Q^b(\t)\}\approx -i\delta^b_a\nonumber\\
&&[\hat{P}_a(\t),\hat{P}_b(\t)]=-i\{P_a(\t),P_b(\t)\}\approx 0\nonumber\\
&&[\hat{Q}^a(\t),\hat{Q}^b(\t)]=-i\{Q^a(\t),Q^b(\t)\}\approx 0.\label{CCR}
\ee
Given the quantum algebra $\Fa$, one can find the representation Hilbert space $\ch$ of $\Fa$ via GNS construction by any positive linear functional on $\Fa$. Note that we can call $\ch$ the physical Hilbert space because all the constraints have been solved in the classical level. Furthermore, 

\begin{Definition}
We say that the quantum dynamics exists for the present system provided that there exists a representation $\ch$ such that for each $\t$ all the physical Hamiltonians $H_I(Q^a(\t),P_a(\t),\t)$ are represented as the densely defined self-adjoint operators on $\ch$. We say that the quantum dynamics is anomaly-free provided that all physical Hamiltonians $H_I(Q^a(\t),P_a(\t),\t)$ form a commutative algebra for each $\t$.
\end{Definition}

If the quantum dynamics is anomaly-free, it means that all the classical gauge symmetries are manifestly reproduced in the quantum theory. Suppose we have a representation of $\Fa$ such that an anomaly-free quantum dynamics exists, on this representation, we have the Heisenberg picture ``equation of motion" from Eq.(\ref{Hamiltonian}):
\be
[H_I(\hat{Q}^a(\t),\hat{P}_a(\t),\t), \hat{O}_f(\t)]=-i\{H_I(Q^a(\t),P_a(\t),\t), O_f(\t)\}\approx -i\frac {\partial}{\partial\t^I}\hat{O}_f(\t)\label{quantumHamiltonian}
\ee
thus the operators $\hat{H}_I(\t)\equiv H_I(\hat{Q}^a(\t),\hat{P}_a(\t),\t)$ are the physical Hamiltonian operators of the quantum system, which generate the multi-finger evolutions of the Heisenberg operators. On the other hand, The time-dependent operators $\hat{H}_I(\t):=H_I(\hat{Q}^a(0),\hat{P}_a(0),\t)$ generate the multi-finger evolution of quantum states in Sch\"odinger picture. For each $I$, a unitary propagator $U_I(\t_I,\t_I')$ is defined by a formal Dyson expansion:
\be
U_I(\t_I,\t_I'):=1+\sum_{n=1}^{\infty}(-i)^{n}\int_{\t_I'}^{\t_I}\rmd \t_{I,n}\int_{\t_I'}^{\t_{I,n}}\rmd \t_{I,n-1}\cdots\int_{\t_I'}^{\t_{I,2}}\rmd \t_{I,1}\ \hat{H}_I\big(\t_{I,1}\big)\cdots\hat{H}_I\big(\t_{I,n}\big)
\ee
which at least formally solves the Sch\"odinger equation
\be
i\frac{\rmd}{\rmd \t_I}U_I(\t_I,\t_I')=\hat{H}_I\big(\t_I\big)U_I(\t_I,\t_I').
\ee
Since the representation is anomaly-free, we define the multi-finger evolution unitary propagator $U(\t,\t')$ by
\be
U(\t, \t'):=\prod_{I}U_I(\t_I,\t_I')
\ee
Given two ``multi-finger time" $\t_N=\{\t_{I,N}\}_I,\t_0=\{\t_{I,0}\}_I$ and two physical state (Heisenberg states) $\Psi,\Psi'\in\ch$, their corresponding Sch\"odinger states are denoted by $\Psi(\t_N),\Psi'(\t_0)$. The physical multi-finger transition amplitude is defined by the physical inner product between these two physical Heisenberg states $\lag\Psi|\Psi'\rag$. We then perform the standard skeletonization procedure \cite{brown} for this physical transition amplitude:
\be
&&\lag\Psi|\Psi'\rag\ =\ \lag\Psi(\t_N)|U(\t_N, \t_0)|\Psi'(\t_0)\rag\nonumber\\
&=&\int\rmd Q^a(\t_N)\rmd Q^a(\t_{N-1})\cdots\rmd Q^a(\t_1)\rmd Q^a(\t_0)\nonumber\\
&&\times\lag\Psi|Q^a(\t_N)\rag\lag Q^a(\t_N)|Q^a(\t_{N-1})\rag\cdots\lag Q^a(\t_1)|Q^a(\t_{0})\rag\lag Q^a(\t_0)|\Psi'\rag\label{skeleton}
\ee
where $|Q^a(\t) \rag\lag Q^a(\t)|$ is the projection valued measure associated with $\hat{Q}^a(\t)$ (we assume $\hat{Q}^a(\t)$ is represented as a self-adjoint operator for each $\t$). From Eq.(\ref{skeleton}), we see that a (discrete) path $c$ in the space of $\t$ is selected for this skeletonization precedure. We denote by $\ct$ the space of $\t$ and by $c:\mathbb{R}\to\ct$ the path parametrized by the parameter $t$, and $c(t_n)=\t_n$. We will call this parameter $t$ the ``external time parameter". However the value of $\lag\Psi|\Psi'\rag$ is manifestly independent of the choice of the path $c$ (external-time reparametrization) by the anomaly-freeness. This fact is a reflection of the general covariance of the system. 

Following the way in \cite{brown}, we arrive at a formal path-integral formula of the physical inner product of Heisenberg states:
\be
&&\lag\Psi|\Psi'\rag\nonumber\\
&=&\int \Big[\prod_{n=0}^{N-1}\rmd P_a(c(t_n))\Big]\Big[\prod_{n=0}^N\rmd Q^a(c(t_n))\Big]\ \overline{\Psi\Big(Q^a(c(t_N)),c(t_N)\Big)}\ {\Psi'}\Big(Q^a(c(t_0)),c(t_0)\Big)\nonumber\\
&&\times \exp i\sum_{n=1}^N\Big[\sum_aP_a(c(t_{n-1}))\Big(Q^a(c(t_n))-Q^a(c(t_{n-1}))\Big)-\sum_I\Big(c^I(t_{n})-c^I(t_{n-1})\Big)H_I(c(t_{n-1}))\Big]
\ee
Formally take the continuous limit $N\to\infty$, we obtain a formal Hamiltonian path-integral expression:
\be
&&\lag\Psi|\Psi'\rag\nonumber\\
&=&\int\prod_{t\in[t_i,t_f]}\Big[\rmd P_a(c(t))\ \rmd Q_a(c(t))\Big]\ e^{i\int_{t_i}^{t_f}\rmd t\lt[P_a(c(t))\dot{Q}_a(c(t))-{H}_I(c(t))\dot{c}^I(t)\rt]}\ \overline{\Psi\Big(Q^a(c(t_f)),c(t_f)\Big)}\ {\Psi'}\Big(Q^a(c(t_i)),c(t_i)\Big)\nonumber\\
&=&\int\prod_{t\in[t_i,t_f]}\Big[\rmd P_a(c(t))\ \rmd Q_a(c(t))\ \rmd P_I(t)\ \rmd T^I(t)\Big]\ \prod_{t\in[t_i,t_f]}\delta\Big(P_I(t)+H_I\big(c(t)\big)\Big)\ \delta\lt(T^I(t)-c^I(t)\rt)
e^{i\int_{t_i}^{t_f}\rmd t\lt[P_a(c(t))\dot{Q}_a(c(t))+{P}_I(t)\dot{T}^I(t)\rt]}\nonumber\\ 
&&\times\overline{\Psi\Big(Q^a(c(t_f)),T^I(t_f)\Big)}\ {\Psi'}\Big(Q^a(c(t_i)),T^I(t_i)\Big)
\ee
Because of the $\delta$-functions $\delta\Big(T^I(t)-c^I(t)\Big)$, each $Q^a(\t)\approx[\a_\b(q^a)]_{\b^I=\t^I-T^I}$ reduces to $q^a$ and the same for the momenta $P^a(\t)$. Therefore
\be
&&\lag\Psi|\Psi'\rag\nonumber\\
&=&\int\prod_{t\in[t_i,t_f]}\Big[\rmd p_a(t)\ \rmd q_a(t)\ \rmd P_I(t)\ \rmd T^I(t)\Big]\ \prod_{t\in[t_i,t_f]}\delta\Big(P_I(t)+H_I(t)\Big)\ \delta\lt(T^I(t)-c^I(t)\rt)
e^{i\int_{t_i}^{t_f}\rmd t\lt[p_a(t)\dot{q}_a(t)+{P}_I(t)\dot{T}^I(t)\rt]}\nonumber\\ 
&&\times\overline{\Psi\Big(q^a(t_f),T^I(t_f)\Big)}\ {\Psi'}\Big(q^a(t_i),T^I(t_i)\Big)\label{RPI1}
\ee  
Eq.(\ref{RPI1}) can also be shortly written as
\be
&&\lag\Psi|\Psi'\rag\nonumber\\
&=&\int \cd p_a\cd q^a\cd P_I\cd T^I\ \prod_{t,I}\left[\delta\Big(P_I+h_I\Big)\ \delta\Big(T^I-\t^I\Big)\right]\ e^{ i\int_{t_i}^{t_f}\rmd t\left[\sum_ap_a(t)\dot{q}^a(t)+\sum_IP_I(t)\dot{T}^I(t) \right]}\ \overline{\Psi(q^a_f,T^I_f)}\ {\Psi'}(q^a_i,T^I_i)\label{RPI2}
\ee
This result implies that the path-integral amplitude with boundary \emph{kinematical states} ${\Psi(q^a_f,T^I_f)}$ and ${\Psi'}(q^a_i,T^I_i)$ can be interpreted as the physical inner product between the corresponding \emph{physical states} in the relational framework (we will come back and discuss more about this point in section \ref{PIARM}). On the other hand, it is also a Faddeev-Popov path-integral formula with the gauge fixing functions $T^I-\t^I$ and unit Faddeev-Popov determinant. We can also obtain the path-integral with the original constraint $C_I$ and general gauge fixing condition $\xi^I$ (two sets of the equations $\xi^I=0$ and $T^I=\t^I$ have to share the same solutions) by the relation:
\be
\sqrt{|D_1|}\prod_I\left[\delta(C_I)\ \delta(\xi^I)\right]&\equiv&\det\left(\{C_I,\xi^J\}\right)\ \prod_I\left[\delta(C_I)\ \delta(\xi^I)\right]\ =\ \det\left(\{P_K,T^L\}\right)\ \det\left(\frac{\partial C_I}{\partial P_K}\right)\ \det\left(\frac{\partial \xi^J}{\partial T^L}\right)\prod_I\left[\delta(C_I)\ \delta(\xi^I)\right]\nonumber\\
&=&\det\left(\{P_I,T^J\}\right)\ \prod_I\left[\delta(P_I+h_I)\ \delta(T^I-\t^I)\right]
\ee
where $\sqrt{|D_1|}$ denotes the Faddeev-Popov determinant. Recall that if the dynamical system also has a number of second-class constraints $\Phi_i$, the formal measure $\cd p_a\cd q^a\cd P_I\cd T^I$ is the Liouville measure on the constraint surface where all second-class constraints vanish. Then it turns out that (the proof will be shown shortly later) 
\be
\cd p_a\cd q^a\cd P_I\cd T^I=\cd x^A(t) \prod_{t\in[t_i,t_f]}\sqrt{\det \o[x^A(t)]}\ \sqrt{|D_2[x^A(t)]|}\ \delta\Big(\Phi_i[x^A(t)]\Big)
\ee
where $\cd x^A(t) \prod_{t}\sqrt{\det \o[x^A(t)]}$ is the Liouville measure on the full phase space, $\Phi_\a$ are the second-class constraints, and $|D_2|=\det\left(\{\Phi_\a, \Phi_\b\}\right)$ is the Dirac determinant. Inserting these relations into Eq.(\ref{RPI2})
\be
&&\lag\Psi|\Psi'\rag\nonumber\\
&=&\int \cd x^A(t)\prod_{t\in[t_i,t_f]}\sqrt{\det \o[x^A(t)]}\prod_{t\in[t_i,t_f]}\left[\sqrt{|D_1[x^A(t)]|}\ \delta\Big(C_I[x^A(t)]\Big)\ \delta\Big(\xi^I[x^A(t)]\Big)\right]\prod_{t\in[t_i,t_f]} \left[\sqrt{|D_2[x^A(t)]|}\ \delta\Big(\Phi_i[x^A(t)]\Big)\right]\nonumber\\
&&\times \exp\left( iS[x^A(t)]\right)\ \overline{\Psi[x^A(t_f)]}\ {\Psi'}[x^A(t_i)]\ \ \ \ \ \ \ \ \ \ \label{RPI3}
\ee
where $S[x^{A}(t)]$ denotes the action of the system. It is remarkable that this path-integral Eq.(\ref{RPI3}) is independent of the reparametrization of the external time $t$, since the physical inner product (multi-finger physical transition amplitude) is manifestly independent of the choice of path $c$. This fact reflects that we are dealing with a general reparametrization-invariant system, whose Hamiltonian is a linear combination of constraints.

Finally we write down the partition function from relational framework:
\be
\cz_{Relational}
&=&\int \cd x^A(t)\prod_{t\in[t_i,t_f]}\sqrt{\det \o[x^A(t)]}\prod_{t\in[t_i,t_f]}\left[\sqrt{|D_1[x^A(t)]|}\ \delta\Big(C_I[x^A(t)]\Big)\ \delta\Big(\xi^I[x^A(t)]\Big)\right]\nonumber\\
&&\times \prod_{t\in[t_i,t_f]} \left[\sqrt{|D_2[x^A(t)]|}\ \delta\Big(\Phi_i[x^A(t)]\Big)\right]\ \exp\left( iS[x^A(t)]\right).\label{RPI4}
\ee

I will show in the follows that the path-integral formula Eq.(\ref{RPI4}) coincides with the reduced phase space path-integral quantization using Liouville measure \cite{QuantizationGauge}. The following discussion also (1) includes the case that the Hamiltonian is non-vanishing on the constraint surface; (2) manifestly shows that different canonical formulations of the same constrained system result in the equivalent path-integral quantizations. 

We consider a general regular dynamical system associated with a $2n$-dimensional phase space $(\G,\O)$ with symplectic coordinates $(p_a,q^a)$ ($a=1,\cdots,n$). There are $m$ first-class constraints $C_I$ ($I=1,\cdots,m$) and $2k$ second-class constraint $\Psi_i$ ($i=1,\cdots,2k$) for this dynamical system, and we suppose $m+k<n$ in order that there are still some unconstrained dynamical degree of freedom. Note that since the constraints $C_I$ are first-class, they close under Poisson bracket determined by $\O$ modulo second-class constraints, and weakly Poisson commute with the second-class constraints $\Psi_i$, i.e.
\be
\left\{C_I,C_J\right\}=f_{IJ}^{\ \ K}C_K+g_{IJ}^{\ \ k}\Psi_k\ \ \ \ \ \ \mathrm{and}\ \ \ \ \ \ \left\{C_I,\Psi_i\right\}=u_{Ii}^{\ \ J}C_J+v_{Ii}^{\ \ j}\Psi_j
\ee
where $f_{IJ}^{\ \ K}$, $g_{IJ}^{\ \ k}$, $u_{Ii}^{\ \ J}$ and $v_{Ii}^{\ \ j}$ are in general the structure functions depending on the phase space coordinates.
 
Now the total Hamiltonian $H_{tot}$ is written as:
\be
H_{tot}=H+f^i\Psi_i+\l^IC_I
\ee
where $\l^I$ are the free Lagrangian multipliers but $f^i$ are determined as functions on the constraint surface by the consistency of all the constraints. $H$ is the non-vanishing Hamiltonian on the reduced phase space, which is a function weakly commute with the first class constraints, i.e
\be
\left\{H,C_I\right\}=U_I^JC_J+V_I^j\Psi_j
\ee
where $U_I^J$ and $V_I^j$ are in general the functions depending on the phase space coordinates.

In order to have a unique equation of motion without the free Lagrangian multipliers, one can introduce $m$ gauge fixing functions $\chi^I$ ($C_I$ and $\chi^I$ shouldn't be weakly Poisson commute) to locally cut the gauge obits generated by the first class constraints (we ignore the potential problems about Gribov copies). The implementation of the local gauge fixing conditions $\chi^I\approx0$ reduce the original first-second-class-mixed constrained system into a purely second-class constrained system. Thus the total Hamiltonian is re-defined by
\be
H_{tot}=H+f^i\Psi_i+\l^IC_I+\b_I\chi^I
\ee  
By the consistency of $C_I$ and $\chi^I$, their coefficients $\l^I$ and $\b_I$ are determined as some certain functions on the constraint surface defined by $\Psi_i=C_I=\chi^I=0$. This constraint surface $\G_R$ combined with the Dirac symplectic structure $\O_R$ determined by the constraints $C_I$, $\chi^I$, and $\Psi_i$ is (at least locally) symplectic isomorphic to the reduced phase space for the constrained system (see \cite{QuantizationGauge} for proof). 

Then we consider the path-integral quantization of a general regular\footnote{A regular constrained system means that its Dirac matrix has a constant rank on the phase space.} dynamical system with both the first-class and the second-class constraints. We have shown that this kind of constrained system can always be reduced into a purely second-class constrained system by introducing a certain number of gauge fixing conditions, so it is enough to consider the path-integral quantization for a general regular dynamical system only associated with the second-class constraints \cite{QuantizationGauge}. We denote its $2n$-dimensional phase space by $(\Gamma,\Omega)$ with a general coordinates $x^I$ $I=1,\cdots,2n$. Suppose there is a irreducible set of $2m$ regular second-class constraints $\chi_\alpha$ $\alpha=1,\cdots,2m$, then the reduced phase space $(\Gamma_R,\Omega_R)$ is defined by the sub-manifold $\G_R=\{x\in\Gamma:\chi_\alpha(x)=0,\ \forall\alpha=1,\cdots,2m\}$ with the Dirac symplectic structure $\omega_R$ determined by $\chi_\a$. On the reduced phase space $(\Gamma_R,\Omega_R)$, all the degree of freedom is free of constraint, thus it is straight-forward to (heuristically) define the path-integral partition function. We introduce $2(n-m)$ coordinates $y^i$ $i=1,\cdots,2(n-m)$ on $\G_R$, then the partition function of the system is defined by a path-integral with respect to an infinite product of Liouville measure associated to $\omega_R$:
\begin{eqnarray}
Z:=\int\prod_{t\in[t_1,t_2]}\left[\rmd y^{i}(t)\sqrt{\left|\det\omega_R[y^i(t)]\right|}\ \right]\exp(iS[y^i(t)])
\end{eqnarray}
where $S[y^{i}(t)]$ denotes the action of the system.

In order to rewrite the path-integral formula in terms of the original phase space coordinates $x^I$. We make a coordinate transformation on $\Gamma$ from $\{x^I\}_{I=1}^{2n}$ to $\{y^i\}_{i=1}^{2(n-m)}$ and $\{\chi_\alpha\}_{\alpha=1}^{2m}$, where $\{y^{i}\}_{i=1}^{2(n-m)}$ are the coordinates on $\G_R$ such that $\{y^i,\chi_\alpha\}\approx0$ (it is always possible, see Theorem 2.5 of \cite{QuantizationGauge}). Then the simplectic structure on $\Gamma$ can be written as 
\begin{eqnarray}
\omega=(\omega_\chi)^{\alpha\beta}\mathrm{d}\chi_\alpha\wedge\mathrm{d}\chi_\beta+(\omega_R)_{ij}\mathrm{d}y^i\wedge\mathrm{d}y^j.
\end{eqnarray}
In fact, $(\omega_\chi)^{\alpha\beta}$ is the inverse of the Dirac matrix $\Delta_{\alpha\beta}=\{\chi_\alpha,\chi_\beta\}$ Thus we obtain a relation of the integration measure by the invariance of the Liouville measure under the coordinate transformation:
\begin{eqnarray}
\sqrt{\left|\det\omega[x^I]\right|}\prod_I\mathrm{d}x^I=\sqrt{\left|\frac{\det\omega_R[y^i]}{\det\Delta}\right|}\prod_\alpha\mathrm{d}\chi_\alpha\prod_i\mathrm{d}y^i.\nonumber
\end{eqnarray}
Therefore we obtain the desired path-integral formula in terms of $x^I$
\begin{eqnarray}
Z=\int\prod_{t\in[t_1,t_2]}\left[\rmd x^{I}(t)\sqrt{\left|\det\omega[x^I(t)]\right|}\ \right]\prod_{t\in[t_1,t_2]}\left[\sqrt{\left|\det\Delta[x^I(t)]\right|}\prod_\alpha\delta(\chi_\alpha[x^I(t)])\right]\exp(iS[x^I(t)])\label{generalZ}
\end{eqnarray}
Note that since this path-integral formula is a quantization on the reduced phase space, the partition function $Z$ is independent of the choice of the gauge fixing function $\chi_\a$.

We apply the path-integral formula Eq.(\ref{generalZ}) to our general first-second-class-mixed constraint system. The phase space coordinates $x^I$ are chosen to be the symplectic coordinates $(p_a,q^a)$. The second-class constraints $\chi_\a$ are chosen to be the first-class constraints $C_I$, the gauge fixing functions $\chi^I$, and the second-class constraints $\Psi_i$. On the constraint surface, the Dirac matrix $\Delta$ takes the following form:
\begin{center}
\begin{tabular}{|c||c|c|c|}
\hline
$\Delta_{\a\b}$ & $C_I$ & $\Psi_i$ & $\chi^I$ \\
\hline\hline
$C_J$ & 0& 0 &$\Delta_{FP}$ \\
\hline
$\Psi_j$ &0&$\Delta_D$&$\Theta$\\
\hline
$\chi^J$& $-\Delta_{FP}$&$-\Theta$ & $\Xi$\\
\hline
\end{tabular}
\end{center}
where $(\Delta_D)_{ij}:=\{\Psi_i,\Psi_j\}$ is the Dirac matrix for the second-class constraint $\Psi_i$, and $(\Delta_{FP})_{IJ}:=\{C_I,\chi^J\}$ is the Faddeev-Popov matrix for the gauge fixing functions $\chi^I$. The absolute value for the determinant of $\Delta$ is
\be
\left|\det\Delta\right|=\left|\det\Delta_D\right|\left|\det\Delta_{FP}\right|^2
\ee
Therefore, the partition function of the constrained system is written as
\be
Z&=&\int\prod_{t\in[t_1,t_2]}\left[\rmd p_a(t)\rmd q^a(t)\sqrt{\left|\det\Delta_D\left[p_a(t), q^a(t)\right]\right|}\prod_{i=1}^{2k} \delta \Big(\Psi_i\left[p_a(t), q^a(t)\right] \Big)\right]\prod_{t\in[t_1,t_2]}\left[\prod_{I=1}^m\delta\Big(C_I\left[p_a(t), q^a(t)\right]\Big)\right]\nonumber\\
&&\times\prod_{t\in[t_1,t_2]}\left[\Big|\det\Delta_{FP}\left[p_a(t), q^a(t)\right]\Big|\prod_{I=1}^m\delta\Big(\chi^I\left[p_a(t), q^a(t)\right]\Big)\right]\exp\Big(iS\left[p_a(t), q^a(t)\right]\Big)
\ee
which precisely coincides with the path-integral formula Eq.(\ref{RPI4}) from the relational framework and the canonical quantization on the reduced phase space, when the Hamiltonian $H$ vanishes. Here we have split the integrand into four different factors in order to clarify the different physical meanings. 

\begin{enumerate}
\item The first factor
\be
\prod_{t\in[t_1,t_2]}\left[\rmd p_a(t)\rmd q^a(t)\sqrt{\left|\det\Delta_D\left[p_a(t), q^a(t)\right]\right|}\prod_{i=1}^{2k} \delta \Big(\Psi_i\left[p_a(t), q^a(t)\right] \Big)\right]
\ee
is essentially an infinite product of the Liouville measures on the constraint surface $\G_\Psi:=\left\{x\in\G\ |\ \Psi_i\left[x\right]=0\right\}$ equipped with the Dirac symplectic structure $\O_\Psi$. This product Liouville measure is invariant under the canonical transformations on $(\G_\Psi,\O_\Psi$. 

\item The second factor 
\be
\prod_{t\in[t_1,t_2]}\left[\prod_{I=1}^m\delta\Big(C_I\left[p_a(t), q^a(t)\right]\Big)\right]
\ee
is a product of the $\delta$-functions of the first-class constraints $C_I$, which will be exponentiated and contribute the total Hamiltonian, after we use the Fourier decompositions of the $\delta$-functions. The third factor 
\be
\prod_{t\in[t_1,t_2]}\left[\Big|\det\Delta_{FP}\left[p_a(t), q^a(t)\right]\Big|\prod_{I=1}^m\delta\Big(\chi^I\left[p_a(t), q^a(t)\right]\Big)\right]
\ee
is a typical Faddeev-Popov term which always appears in the gauge fixed path-integral. 

\item And the last factor is the exponentiated action where the action $S\left[p_a(t), q^a(t)\right]$ reads:
\be
S\left[p_a(t), q^a(t)\right]=\int_{t_1}^{t_2}\rmd t\left[p_a(t)\frac{d}{dt}q^a(t)+H\left[p_a(t), q^a(t)\right]\right].
\ee 

\end{enumerate}
The next step is to employ the Fourier decomposition of the $\delta$-functions so that we write the partition function as
\be
Z&=&\int\prod_{t\in[t_1,t_2]}\left[\rmd p_a(t)\rmd q^a(t)\sqrt{\left|\det\Delta_D\left[p_a(t), q^a(t)\right]\right|}\prod_{i=1}^{2k} \delta \Big(\Psi_i\left[p_a(t), q^a(t)\right] \Big)\right]\prod_{t\in[t_1,t_2]}\left[\prod_{I=1}^m \rmd \l^I(t)\right]\nonumber\\
&&\times\prod_{t\in[t_1,t_2]}\left[\Big|\det\Delta_{FP}\left[p_a(t), q^a(t)\right]\Big|\prod_{I=1}^m\delta\Big(\chi^I\left[p_a(t), q^a(t)\right]\Big)\right]\nonumber\\
&&\times\exp i\int_{t_1}^{t_2}\rmd t\left[p_a(t)\frac{d}{dt}q^a(t)+H\left[p_a(t), q^a(t)\right]+\l^I(t)C_I\left[p_a(t), q^a(t)\right]\right] 
\ee
where we have exponentiated the first-class constraints $C_I$, and replaced the second factor by a formal product measure of the Lagrangian multipliers.

\subsection{The gauge invariance of path-integral measure}

In this subsection, we will consider the invariance of the total measure factor 
\be
D\mu:=\prod_{t\in[t_1,t_2]}\left[\rmd p_a(t)\rmd q^a(t)\sqrt{\left|\det\Delta_D\left[p_a(t), q^a(t)\right]\right|}\prod_{i=1}^{2k} \delta \Big(\Psi_i\left[p_a(t), q^a(t)\right] \Big)\right]\prod_{t\in[t_1,t_2]}\left[\prod_{I=1}^m \rmd \l^I(t)\right]\label{Dmu}
\ee
under the gauge transformations generated by the first-class constraints.

We first consider the first-class constraints reduced on the second-class constraint surface $(\G_\Psi,\O_\Psi)$, we denote the restricted first-class constraints $C_I\big|_{G_{\Psi}}$ also by $C_I$, which won't result in any mis-understanding, because we will restrict all the following discussion on the phase space $(\G_\Psi,\O_\Psi)$. On the phase space $(\G_\Psi,\O_\Psi)$, the constraint algebra is the Poisson algebra generated by $C_I$ under the Dirac bracket $\{,\}_\Psi$ determined by $\O_\Psi$ 
\be
\left\{C_I,C_J\right\}_\Psi&=&\left\{C_I,C_J\right\}-\left\{C_I,\Psi_i\right\}\left(\Delta_D^{-1}\right)^{ik}\left\{\Psi_k,C_J\right\}\nonumber\\
&=&f_{IJ}^{\ \ K}C_K+g_{IJ}^{\ \ k}\Psi_k+\left[u_{Ii}^{\ \ K}C_K+v_{Ii}^{\ \ j}\Psi_j\right]\left(\Delta_D^{-1}\right)^{ik}\left[u_{Jk}^{\ \ L}C_L+v_{Jk}^{\ \ l}\Psi_l\right]\nonumber\\
&\peq&\left[f_{IJ}^{\ \ K}+u_{Ii}^{\ \ L}C_L\left(\Delta_D^{-1}\right)^{ik}u_{Jk}^{\ \ K}\right]C_K\nonumber\\
&\peq&F_{IJ}^{\ \ K}C_K
\ee
where `` $^\Psi\!\!\!\!\!\!=$ " means the equality on $(\G_\Psi,\O_\Psi)$ and $F_{IJ}^{\ \ K}\equiv\left[f_{IJ}^{\ \ K}+u_{Ii}^{\ \ L}C_L\left(\Delta_D^{-1}\right)^{ik}u_{Jk}^{\ \ K}\right]$ is the structure function for the first-class constraint algebra on $(\G_\Psi,\O_\Psi)$. For the Dirac bracket between $C_I$ and the Hamiltonian $H$, we have
\be
\left\{H,C_I\right\}_\Psi&=&\left\{H,C_I\right\}-\left\{H,\Psi_i\right\}\left(\Delta_D^{-1}\right)^{ik}\left\{\Psi_k,C_I\right\}\nonumber\\
&=&U_I^JC_J+V_I^j\Psi_j-\left\{H,\Psi_i\right\}\left(\Delta_D^{-1}\right)^{ik}\left[u_{Ik}^{\ \ L}C_L+v_{Ik}^{\ \ l}\Psi_l\right]\nonumber\\
&\peq&\left[U_I^J-\left\{H,\Psi_i\right\}\left(\Delta_D^{-1}\right)^{ik}u_{Ik}^{\ \ J}\right]C_J\nonumber\\
&\peq&W_I^JC_J
\ee
Now we consider the infinitesimal gauge transformation generated by a first-class constraint $C_J$. For any function $f$ on the phase space $(\G_\Psi,\O_\Psi)$, it is defined by 
\be
f\mapsto \tilde{f}:=f+\eps\left\{f,C_J\right\}_\Psi
\ee
In the path-integral measure Eq.(\ref{Dmu}), there is a factor of the product measure of the Lagrangian multipliers $\left[\prod_{I=1}^m \rmd \l^I(t)\right]$ whose gauge transformation is not obvious so far, because the constraints $C_I$ are the phase space functions and their Hamiltonian vector fields don't have action on the Lagrangian multipliers. However the gauge transformations for the Lagrangian multipliers $\l^I$ is obtained if we ask the action is invariant under gauge transformation generated by the first-class constraints, i.e. the canonical variables and Lagrangian multipliers should transform at the same time and form the local symmetries of the action.

Apply this infinitesimal gauge transformation generated by $C_J$ to the total action $S_{tot}$ (also assume the change of $\l^I$), we obtain the transformed total action by
\be
S_{tot}\left[\tilde{p}_a(t), \tilde{q}^a(t),\tilde{\l}^I(t)\right]&=&\int_{t_1}^{t_2}\rmd t\left[\tilde{p}_a(t)\frac{d}{dt}\tilde{q}^a(t)+\tilde{H}+\widetilde{\l^K(t){C}_K}\right]_{\G_\Psi}\nonumber\\
&=&\int_{t_1}^{t_2}\rmd t\left[{p}_a(t)\frac{d}{dt}{q}^a(t)+H+\eps W_J^KC_K+\tilde{\l}^K(t)C_K+\eps{\l}^IF_{IJ}^{\ \ K}C_K\right]_{\G_\Psi}+o(\eps^2)
\ee 
where we have used the fact that the kinetic term is unchanged under canonical transformation and
\be
\l^IC_I\mapsto\widetilde{\l^IC_I}:=\tilde{\l}^IC_I+\l\tilde{C}_I+o(\eps^2).
\ee
Obviously, if at the same time we transform the Lagrangian multiplier by 
\be
\l^K\mapsto\tilde{\l}^K:=\l^K-\eps\left[W_J^K+\l^IF_{IJ}^{\ \ K}\right]
\ee
the total action $S_{tot}$ is unchanged up to the order of $\eps^2$
\be
S_{tot}\left[\tilde{p}_a(t), \tilde{q}^a(t),\tilde{\l}^I(t)\right]=S_{tot}\left[{p}_a(t), {q}^a(t),\l^I(t)\right]+o(\eps^2)
\ee
or in another word:
\be
\delta_{J,\eps} S_{tot}\left[{p}_a(t), {q}^a(t),\l^I(t)\right]:=\lim_{\eps\to0}\frac{S_{tot}\left[\tilde{p}_a(t), \tilde{q}^a(t),\tilde{\l}^I(t)\right]-S_{tot}\left[{p}_a(t), {q}^a(t),\l^I(t)\right]}{\eps}=0
\ee
Therefore we give a name for this simultaneous transformation 

\begin{Definition}
The simultaneous infinitesimal transformations for both the phase space functions and the Lagrangian multipliers induced by the first-class constraint $C_J$
\be
f\left[p^a,q^a\right]\mapsto \tilde{f}\left[p^a,q^a\right]:=f\left[p^a,q^a\right]+\eps\left\{f\left[p^a,q^a\right],C_J\left[p^a,q^a\right]\right\}_\Psi\ \ \ \ \ {and}\ \ \ \ \ \l^K\mapsto\tilde{\l}^K:=\l^K-\eps\left[W_J^K+\l^IF_{IJ}^{\ \ K}\right]
\ee
which results in $\delta_{J,\eps} S_{tot}=0$ is called the \textbf{infinitesimal local symmetry} generated by $C_J$.
\end{Definition}

There are two remarks:
\begin{itemize}
\item It is well known that the local symmetries of the action in general result in the constraints in the canonical framework. Given an classical action, there is a projection map from the space of field configurations to its phase space. This projection map usually projects the infinitesimal local symmetry transformations of the action to the same number of infinitesimal phase space gauge transformations generated by the constraints \cite{wald}. With respect to this, the construction of infinitesimal gauge transformations for action from constraints is its reverse procedure. But this reverse procedure sometimes cannot capture all the local symmetry transformations, espectially in general relativity, because some local symmetry transformations may not be projectable, e.g. the field-independent spacetime diffeomorphisms.

\item The reason for only considering infinitesimal transformations is the following: Here we are discussing the most general cases in which the first-class constraint algebra is not a Lie algebra (there are some structure functions). Therefore the collection of gauge transformations doesn't have a group structure and actually, is at most an enveloping algebra of the first-class constraint algebra. A generic element of the enveloping algebra is a finite product of infinitesimal gauge transformations.
\end{itemize}

\begin{Definition}
An infinitesimal local symmetry generated by a first-class constraint is said to be implemented quantum mechanically without anomaly if the total path-integral measure 
\be
D\mu\left[{p}_a(t), {q}^a(t),\l^I(t)\right]:=\prod_{t\in[t_1,t_2]}\left[\rmd p_a(t)\rmd q^a(t)\sqrt{\left|\det\Delta_D\left[p_a(t), q^a(t)\right]\right|}\prod_{i=1}^{2k} \delta \Big(\Psi_i\left[p_a(t), q^a(t)\right] \Big)\right]\prod_{t\in[t_1,t_2]}\left[\prod_{I=1}^m \rmd \l^I(t)\right] 
\ee
is invariant (up to an overall constant) under this local symmetry transformation.
\end{Definition}

Under this gauge transformation, the transformation behavior of the total measure is given by
\be
&&\prod_{t\in[t_1,t_2]}\left[\rmd p_a(t)\rmd q^a(t)\sqrt{\left|\det\Delta_D\left[p_a(t), q^a(t)\right]\right|}\prod_{i=1}^{2k} \delta \Big(\Psi_i\left[p_a(t), q^a(t)\right] \Big)\right]\prod_{t\in[t_1,t_2]}\left[\prod_{I=1}^m \rmd \l^I(t)\right]\nonumber\\
&\mapsto&\prod_{t\in[t_1,t_2]}\left[\rmd \tilde{p}_a(t)\rmd \tilde{q}^a(t)\sqrt{\left|\det\Delta_D\left[\tilde{p}_a(t), \tilde{q}^a(t)\right]\right|}\prod_{i=1}^{2k} \delta \Big(\Psi_i\left[\tilde{p}_a(t), \tilde{q}^a(t)\right] \Big)\right]\prod_{t\in[t_1,t_2]}\left[\prod_{I=1}^m \rmd \tilde{\l}^I(t)\right] \nonumber\\
&=&\prod_{t\in[t_1,t_2]}\left[\rmd p_a(t)\rmd q^a(t)\sqrt{\left|\det\Delta_D\left[p_a(t), q^a(t)\right]\right|}\prod_{i=1}^{2k} \delta \Big(\Psi_i\left[p_a(t), q^a(t)\right] \Big)\right]\prod_{t\in[t_1,t_2]}\left[\prod_{I=1}^m \rmd \l^I(t)\right]\Bigg(1-\eps\ F_{IJ}^{\ \ I}\left[p_a(t), q^a(t)\right]\Bigg)                                      
\ee
where we have used the fact that the first factor of the total measure is an infinite product of the Liouville measures on the constraint surface $\G_\Psi$ equipped with the Dirac symplectic structure $\O_\Psi$. This product Liouville measure is invariant under the canonical transformations on $(\G_\Psi,\O_\Psi$. Therefore we have proven the following result:

\begin{Theorem}\label{invariancecondition}
The local symmetry generated by the first-class constraint $C_J$ is implemented without anomaly in reduced phase space quantization if and only if the trace of the structure function, $F_{IJ}^{\ \ I}$, is a phase space constant.
\end{Theorem}

Note that we only need the invariance of the measure $D\mu$ up to an overall constant because essentially the quantities we are computing is the physical inner product $\lag\Psi|\Psi'\rag$ (see Eq.(\ref{RPI3})). More precisely, we may choice a reference vector in the physical Hilbert space $\ch$, the meaningful quantity is the ratio
\be
\lag\Psi|\Psi'\rag_\O=\frac{\lag\Psi|\Psi'\rag}{\lag\O|\O\rag}
\ee 
which is invariant under a re-scaling of the path-integral measure.

\section{The gauge invariance of the path-integral measure of gravity}\label{GRinv}

\subsection{The path-integral measure in ADM formalism}

We would like to apply our general consideration to gravity. First of all we consider the ADM formalism of the 4-dimensional canonical general relativity. The ADM formalism formulates canonical general relativity as a purely first-class constrained system, whose total Hamiltonian is a linear combination of first-constraints:
\be
H_{tot}:=\int_{\Sig}\rmd^{3}x\ \lt[N^a(x)\ H_a(x)+N(x)\ H(x)\rt]
\ee
where $H_a$ and $H$ are spatial diffeomorphism constraint and Hamiltonian constraint respectively. They are expressed as:
\be
H_a&=&-\frac{2}{\kappa}q_{ac}D_b\ P^{bc}\nonumber\\
H&=&-\sqrt{|\det q|}R-\frac{s}{\sqrt{|\det q|}}\left[q_{ac}q_{bd}-\frac{1}{D-1}q_{ab}q_{cd}\rt]P^{ab}P^{cd}
\ee
The first-class constraint algebra is given by
\be
\left\{H_a(N^a),H_b(M^b)\right\}&=&H_a(-\cl_{\vec{M}}N^a)\nonumber\\
\left\{H(N),H_a(N^a)\right\}&=&H(-\cl_{\vec{N}}N)\nonumber\\
\left\{H(N),H(M)\right\}&=&H_a(q^{ab}[N\partial_bM-M\partial_bN])
\ee
It is not hard to see that the constraint algebra is not a Lie algebra since there is a structure function appearing in the commutator between two Hamiltonian constraints. Therefore the gauge transformations generated by these constraints don't form a group but only can form a enveloping algebra, to which we refer as the Bargmann-Komar Group BK(M) \cite{BK}. This enveloping algebra obtains a group structure when the equations of motion is imposed, and only in this case BK(M)=Diff(M). We will come back to this point in Section \ref{BKvsDiff}.

We apply our previous general consideration to ADM formalism and write down the path-integral partition function in terms of canonical variables $(P^{ab}, q_{ab})$
\be
Z_{ADM}&=&\int\prod_{x\in M}\left[\rmd P^{ab}(x)\ \rmd q_{ab}(x)\right]\ \prod_{x\in M}\left[\delta\Big(H_a(x)\Big)\ \delta\Big(H(x)\Big)\right]\ \exp i\int\rmd t\int\rmd^3x\  P^{ab}\frac{\partial}{\partial t}q_{ab}\nonumber\\
&&\times\prod_{x\in M}\left[\Big|\det\Delta_{FP}(x)\Big|\prod_{\a=1}^4 \delta\Big(\chi^\a\left[P^{ab}(x),q_{ab}(x)\right]\Big)\right]\nonumber\\
&=&\int\prod_{x\in M}\left[\rmd P^{ab}(x)\ \rmd q_{ab}(x)\right]\ \prod_{x\in M}\Big[\rmd N^a(x)\ \rmd N(x)\Big]\ \exp i\int\rmd t\int\rmd^3x\  \lt[P^{ab}\frac{\partial}{\partial t}q_{ab}+N^a(x)\ H_a(x)+N(x)H(x)\rt]\nonumber\\
&&\times\prod_{x\in M}\left[\Big|\det\Delta_{FP}(x)\Big|\prod_{\a=1}^4 \delta\Big(\chi^\a\left[P^{ab}(x),q_{ab}(x)\right]\Big)\right]
\ee
where the path-integral measure and the ADM-action read
\be
D\mu_{ADM}&=&\prod_{x\in M}\left[\rmd P^{ab}(x)\ \rmd q_{ab}(x)\right]\ \prod_{x\in M}\Big[\rmd N^a(x)\ \rmd N(x)\Big]\nonumber\\
S_{ADM}[P^{ab},q_{ab},N^a,N]&=&\int\rmd t\int\rmd^3x\  \lt[P^{ab}\frac{\partial}{\partial t}q_{ab}+N^a(x)\ H_a(x)+N(x)H(x)\rt]
\ee
In the following we need to write down the local symmetries of $S_{ADM}$ generated by spatial diffeomorphism constraint and Hamiltonian constraint. These transformations have to transform not only the phase space variables but also the Lagrangian multipliers which are shift vector and lapse function, in order to keep the ADM-action invariant.

First of all, we consider the infinitesimal gauge transformation $T^\eps_{Diff}$ generated by the spatial diffeomorphism constraint $H_a(\eps^a)$. For a phase space function $f[P^{ab},q_{ab}]$, the transformation is given by
\be
T^\eps_{Diff}f:=f+\lt\{f,H_a(\eps^a)\rt\}
\ee
Then we look at the transformation of the action 
\be
&&S_{ADM}\left[P^{ab},q_{ab},N^a,N\right]=\int\rmd t\int\rmd^3x\  \lt[P^{ab}\frac{\partial}{\partial t}q_{ab}+N^a(x) H_a(x)+N(x)H(x)\rt]\nonumber\\
&\mapsto& S_{ADM}\left[T^{\eps}_{Diff}P^{ab},T^{\eps}_{Diff}q_{ab},T^{\eps}_{Diff}{N}^a,T^{\eps}_{Diff}{N}\right]\nonumber\\
&&=\int\rmd t\int\rmd^3x\Bigg[P^{ab}\frac{\partial}{\partial t}q_{ab}+\Big[T^{\eps}_{Diff}N^a-\cl_{\vec{\eps}}N^a\Big]H_a+\Big[T^{\eps}_{Diff}N-\cl_{\vec{\eps}}N\Big]H\ \Bigg]+o(\eps^2)
\ee
where we have used the fact that the kinetic term is unchanged under the canonical transformations, and 
\be
T^\eps_{Diff}H_a(N^a)&=&H_a(T^\eps_{Diff}N^a)+\lt(T^\eps_{Diff}H_a\rt)(N^a)+o(\eps^2)\nonumber\\
T^\eps_{Diff}H(N)&=&H(T^\eps_{Diff}N)+\lt(T^\eps_{Diff}H\rt)(N)+o(\eps^2)
\ee
If we suppose the following transformations assigned to the lapse function and the shift vector correspondingly:
\be
T^{\eps}_{Diff}N^a&=&N^a+\cl_{\vec{\eps}}N^a\nonumber\\
T^{\eps}_{Diff}N&=&N+\cl_{\vec{\eps}}N\label{ADMDiff}
\ee 
the action $S_{ADM}\left[P^{ab},q_{ab},N^a,N\right]$ is invariant up to $o(\eps^2)$ under the transformation generated by the spatial diffeomorphism constraint $T^{\eps}_{Diff}:\left(P^{ab},q_{ab},N^a,N\right)\mapsto\left(T^{\eps}_{Diff}P^{ab},T^{\eps}_{Diff}q_{ab},T^{\eps}_{Diff}N^a,T^{\eps}_{Diff}N\right)$. Therefore
\be
\delta_{{Diff},\vec{\eps}}S_{ADM}\left[P^{ab},q_{ab},N^a,N\right]:=\lim_{||\vec{\eps}||\to0}\frac{S_{ADM}\left[T^{\eps}_{Diff}P^{ab},T^{\eps}_{Diff}q_{ab},T^{\eps}_{Diff}{N}^a,T^{\eps}_{Diff}{N}\right]-S_{ADM}\left[P^{ab},q_{ab},N^a,N\right]}{||\vec{\eps}||}=0
\ee
If the smearing fields $\eps^a(x)$ are the compact support vector fields, the norm $||\vec{\eps}||$ can be chosen as $\sup_{x\in\Sig}\lt|\eps^a(x)\ \eps^a(x)\rt|^{1/2}$. Thus we have shown that $T^{\eps}_{Diff}:\left(P^{ab},q_{ab},N^a,N\right)\mapsto\left(T^{\eps}_{Diff}P^{ab},T^{\eps}_{Diff}q_{ab},T^{\eps}_{Diff}N^a,T^{\eps}_{Diff}N\right)$ is an infinitesimal local symmetry generated by the spatial diffeomorphism constraint. Furthermore it is manifest that the transformation Jacobian in Eq.(\ref{ADMDiff}) is independent of all the fields $\left(P^{ab},q_{ab},N^a,N\right)$, so it is obvious that the path-integral measure 
\be
D\mu_{ADM}&=&\prod_{x\in M}\left[\rmd P^{ab}(x)\ \rmd q_{ab}(x)\right]\ \prod_{x\in M}\Big[\rmd N^a(x)\ \rmd N(x)\Big]
\ee
is invariant under $T^{\eps}_{Diff}$ up to an overall constant.

Then we consider the infinitesimal gauge transformation $T^{\eps}_H$ generated by the Hamiltonian constraint. For any phase space function $f[P^{ab},q_{ab}]$, the transformation is given by
\be
T^\eps_{H}f:=f+\lt\{f,H(\eps)\rt\}
\ee
Then we look at the transformation of the action 
\be
&&S_{ADM}\left[P^{ab},q_{ab},N^a,N\right]=\int\rmd t\int\rmd^3x\  \lt[P^{ab}\frac{\partial}{\partial t}q_{ab}+N^a(x) H_a(x)+N(x)H(x)\rt]\nonumber\\
&\mapsto& S_{ADM}\left[T^{\eps}_{H}P^{ab},T^{\eps}_{H}q_{ab},T^{\eps}_{H}{N}^a,T^{\eps}_{H}{N}\right]\nonumber\\
&&=\int\rmd t\int\rmd^3x\Bigg[P^{ab}\frac{\partial}{\partial t}q_{ab}+\lt(T^{\eps}_{H}N^a\rt)H_a+\lt(\cl_{\vec{N}}\eps\rt) H+\Big(T^{\eps}_{H}N\Big)H+q^{ab}\Big(N\partial_a\eps-\eps\partial_aN\Big)H_b\ \Bigg]+o(\eps^2)
\ee
where we have used the fact that the kinetic term is unchanged under canonical transformations, and 
\be
T^\eps_{H}H_a(N^a)&=&H_a(T^\eps_{H}N^a)+\lt(T^\eps_{H}H_a\rt)(N^a)+o(\eps^2)\nonumber\\
T^\eps_{H}H(N)&=&H(T^\eps_{H}N)+\lt(T^\eps_{H}H\rt)(N)+o(\eps^2)
\ee
If we suppose the following transformations assigned to the lapse function and shift vector correspondingly:
\be
T^{\eps}_{H}N^a&=&N^a-q^{ab}\lt(N\partial_b\eps-\eps\partial_bN\rt)\nonumber\\
T^{\eps}_{H}N&=&N-\cl_{\vec{N}}\eps\ =\ N-N^a\partial_a\eps\label{ADMH}
\ee 
the action $S_{ADM}\left[P^{ab},q_{ab},N^a,N\right]$ then is invariant up to $o(\eps^2)$ under the gauge transformation generated by the Hamitonian constraint $T^{\eps}_{H}:\left(P^{ab},q_{ab},N^a,N\right)\mapsto\left(T^{\eps}_{H}P^{ab},T^{\eps}_{H}q_{ab},T^{\eps}_{H}N^a,T^{\eps}_{H}N\right)$. Therefore
\be
\delta_{{H},{\eps}}S_{ADM}\left[P^{ab},q_{ab},N^a,N\right]:=\lim_{||\vec{\eps}||\to0}\frac{S_{ADM}\left[T^{\eps}_{H}P^{ab},T^{\eps}_{H}q_{ab},T^{\eps}_{H}{N}^a,T^{\eps}_{H}{N}\right]-S_{ADM}\left[P^{ab},q_{ab},N^a,N\right]}{||{\eps}||}=0
\ee
If the smearing fields $\eps(x)$ are compact support functions, the norm $||{\eps}||$ can be chosen as $\sup_{x\in\Sig}\lt|\eps(x)\rt|$. Thus we have shown that $T^{\eps}_{H}:\left(P^{ab},q_{ab},N^a,N\right)\mapsto\left(T^{\eps}_{H}P^{ab},T^{\eps}_{H}q_{ab},T^{\eps}_{H}N^a,T^{\eps}_{H}N\right)$ is an infinitesimal local symmetry generated by the Hamiltonian constraint. Furthermore the transformation Jacobian $\partial \left(T^\eps_{H}N^a,T^\eps_{H}N\right)/\partial\left(N^a,N\right)$ in Eq.(\ref{ADMH}) is expressed as 
\begin{center}
\begin{tabular}{|c||c|c|}
\hline
$\frac{\partial \left(T^\eps_{H}N^a,T^\eps_{H}N\right)}{\partial\left(N^a,N\right)}$ & $T^\eps_{H}N^a$ & $T^\eps_{H}N$ \\
\hline
\hline
$N^a$ & $1$ & $-\partial_a\eps$\\
\hline
$N$& $q^{ab}\left(\eps\partial_b-\partial_b\eps\right)$ & $1$\\
\hline
\end{tabular}$\ =\ I+\ce$
\end{center}
where the perturbation matrix $\ce$ is traceless. Thus the Jacobian determinant is $\det\lt(I+\ce\rt)=1+\mathrm{Tr}\ce=1$. As a result the path-integral measure 
\be
D\mu_{ADM}&=&\prod_{x\in M}\left[\rmd P^{ab}(x)\ \rmd q_{ab}(x)\right]\ \prod_{x\in M}\Big[\rmd N^a(x)\ \rmd N(x)\Big]
\ee
is invariant under $T^{\eps}_{H}$.

So far we assumed the infinitesimal transformation parameter $\eps^\mu=(\eps,\vec{\eps})$ doesn't depend on the time-parameter $t$. However if we let $\eps^\mu=\eps^\mu(t,\vec{x})$, the Eqs.(\ref{ADMDiff}) and (\ref{ADMH}) have generalized expressions. Now we write Eqs.(\ref{ADMDiff}) and (\ref{ADMH}) in a uniform expression and add certain time-derivative terms
\be
T^{\eps^\mu}N^a&=&N^a+\cl_{\vec{\eps}}N^a+q^{ab}(\eps\partial_bN-N\partial_b\eps)+\dot{\eps}^a\nonumber\\
T^{\eps^\mu}N&=&N+\eps^a\partial_aN-N^a\partial_a\eps+\dot{\eps}\label{ADMgen}
\ee
Firstly, combined with the gauge transformations for phase space variables, one can check that Eq.(\ref{ADMgen}) leaves both the action $S_{ADM}$ and the path-integral measure $D\mu_{ADM}$ invariant (the consistency conditions $\dot{H}=\dot{H}_a=0$ should be used to show the invariance of the action). Secondly the generalization Eq.(\ref{ADMgen}) is motivated by the spacetime diffeomorphisms (see section \ref{BKvsDiff}) and can be derived systematically \cite{Pons}. In order to derive it, the phase space has to be enlarged to include the lapse $N$ and the shift $N^a$ as well as their conjugated momenta $\Pi$ and $\Pi_a$. The set of the first-class constraints should also be extended to include $\Pi\approx0$ and $\Pi_a\approx0$. Then the extended gauge transformation generator
\be
H(\eps)+H_a(\eps^a)+\int\rmd^3x\lt[\Pi_a\lt(\cl_{\vec{\eps}}N^a+q^{ab}(\eps\partial_bN-N\partial_b\eps)+\dot{\eps}^a\rt)+\Pi\lt(\eps^a\partial_aN-N^a\partial_a\eps+\dot{\eps}\rt)\rt]
\ee
generates the gauge transformations on all ten components of the spacetime metric, which coincide with spacetime diffeomorphism if the equations of motion are imposed.

Let's summarize all above results:
\begin{Theorem}\label{ADMinvmeasure}
There exists a collection of infinitesimal transformations $T^{\eps^\mu}$ generated by the spatial diffeomorphism constraint $H_a$, and the Hamiltonian constraint $H$:
\be
f\left[P^{ab},q_{ab}\right]&\mapsto& T^{\eps^\mu}f\left[P^{ab},q_{ab}\right]:=f\left[P^{ab},q_{ab}\right]+\left\{f\left[P^{ab},q_{ab}\right],H(\eps)+H_a(\eps^{a})\right\}\nonumber\\
N^a&\mapsto&T^{\eps^\mu}N^a:=N^a+\cl_{\vec{\eps}}N^a+q^{ab}(\eps\partial_bN-N\partial_b\eps)+\dot{\eps}^a\nonumber\\
N&\mapsto&T^{\eps^\mu}N:=N+\eps^a\partial_aN-N^a\partial_a\eps+\dot{\eps}
\ee
These transformations are the infinitesimal local symmetries of the action, i.e. 
\be
\delta_{{\eps^\mu}}S_{ADM}=\lim_{||\eps^\mu||\to0}\frac{S_{ADM}\left[T^{\eps^\mu}P^{ab},T^{\eps^\mu}q_{ab},T^{\eps^\mu}N^a,T^{\eps^\mu}N\right]-S_{ADM}\left[P^{ab},q_{ab},N^a,N\right]}{||\eps^\mu||}=0
\ee
for $\eps^{\mu}$ compact support and $||\eps^\mu||=\sup_{(t,\vec{x})\in M}|\eps^\mu(t,\vec{x})\eps^\mu(t,\vec{x})|^{1/2}$. And the total path-integral measure
\be
D\mu_{ADM}&=&\prod_{x\in M}\left[\rmd P^{ab}(x)\ \rmd q_{ab}(x)\right]\ \prod_{x\in M}\Big[\rmd N^a(x)\ \rmd N(x)\Big]
\ee
is invariant under these transformations (up to an overall constant).
\end{Theorem}

\begin{Definition}
The Bergmann-Komar group BK(M) is the enveloping algebra generated by the collection of $T^{\eps^\mu}$ as the transformations on the space of metric $g_{\a\b}=(q_{ab},N,N^a)$, i.e. 
\be 
\text{BK(M)}:=\cf(\{T^{\eps^\mu}\}_{\eps^\mu})/\sim
\ee
$\{T^{\eps^\mu}\}_{\eps^\mu}$ is the free algebra generated by $T^{\eps^\mu}$ whose generic element read
\be
T^{\eps_1^\mu}T^{\eps_2^\mu}\cdots T^{\eps_n^\mu}
\ee
and ``$\sim$" denotes the equivalence relations from the linearity and the commutation relations of $T^{\eps^\mu}$.
\end{Definition}

\begin{Corollary}
The Bergmann-Komar group BK(M) is a collection of the infinitesimal local symmetries of classical Einstein-Hilbert action, which are implemented without anomaly in the reduced phase space path-integral quantization.
\end{Corollary}

\subsection{The field-independent spacetime diffeomorphism group}\label{BKvsDiff}

There exists another collection of the local symmetries of the Einstein-Hilbert action, which is the set of the \emph{field-independent} spacetime diffeomorphisms Diff(M). The generators of Diff(M) are Lie-derivatives $\cl_u$ along the 4-vector field $u^\mu$. The commutator between two Lie-derivatives 
\be
\lt[\cl_u,\cl_{u'}\rt]=\cl_{[u,u']}
\ee 
endows the collection of generators a Lie algebra structure thus gives Diff(M) a group structure. In order to compare the the \emph{field-independent} spacetime diffeomorphisms and the local symmetries in Bergamnn-Komar group and check if the local symmetries in Diff(M) can be implemented without anomaly in the above path-integral quantization, we have to show how the diffeomorphisms in Diff(M) act on the components $(q_{ab},N,N^a)$ (for details see \cite{GHTW} and the references therein).

The components $(q_{ab},N,N^a)$ can be solved from 4-metric $g_{\a\b}$ by the following relations
\be
N^a=g^{ab}g_{tb},\ \ \ \ \ N^2=-g_{tt}+g^{ab}g_{ta}g_{tb}, \ \ \ \ \ q_{ab}=g_{ab}
\ee
where $g^{ab}$ is the inverse of the spatial metric $g_{ab}$. The infinitesimal change of the 4-metric under the spacetime diffeomorphism is given by the Lie-derivative
\be
\delta_ug_{\a\b}:=\cl_ug_{\a\b}=u^\mu \partial_\mu g_{\a\b}+2g_{\mu(\a}\partial_{\b)}u^\mu
\ee 
To derive the induced transformation for lapse and shift, one should use the relation $u^\mu=\eps n^\mu+X^\mu_{,a}\eps^a$ or 
\be
u^t=\frac{\eps}{N},\ \ \ \ \ u^a=\eps^a-\eps\frac{N^a}{N}
\ee
After some calculus, the infinitesimal changes of lapse and shift under spacetime diffeomorphism is obtained explicitly
\be
\delta_uN^a&=&\cl_{\vec{\eps}}N^a+q^{ab}(\eps\partial_bN-N\partial_b\eps)+\dot{\eps}^a\nonumber\\
\delta_uN&=&\eps^a\partial_aN-N^a\partial_a\eps+\dot{\eps}
\ee  
which coincides with Eq.(\ref{ADMgen}) and actually is the motivation for adding the time-derivative terms in Eq.(\ref{ADMgen}). On the other hand, the infinitesimal changes of spatial metric $q_{ab}$ and its momentum $P^{ab}$ are represented as the Lie-derivatives of the spatial fields
\be
\delta_uq_{ab}:=\cl_uq_{ab},\ \ \ \ \ \ \delta_uP^{ab}:=\cl_uP^{ab}
\ee
However, only when equations of motion is imposed,
\be
\cl_uq_{ab}(x)=\left\{q_{ab}(x),H(\eps)+H_a(\eps^{a})\right\}, \ \ \ \ \ \ \cl_uP^{ab}(x)=\left\{P^{ab}(x),H(\eps)+H_a(\eps^{a})\right\}
\ee 
Thus we see that the relation BK(M)=Diff(M) holds only when the equations of motion are imposed.

Then the question is whether the local symmetries in Diff(M) can be implemented quantum mechanically without anomaly. It turns out in \cite{wald,Pons} that all the infinitesimal field-independent non-spatial diffeomorphisms cannot be taken from the space of metric to the phase space by the projection map in \cite{wald}. In another word, there is no notion of the local symmetries on the phase space corresponding to the non-spatial diffeomorphisms. Thus any non-spatial diffeomorphism belonging to Diff(M) cannot be written as a canonical transformation (unless on shell) and doesn't leave the Liouville measure $\prod_{x\in M}\left[\rmd P^{ab}(x)\ \rmd q_{ab}(x)\right]$ invariant. In conclusion, the classical local symmetries in the spacetime diffeomorphism group Diff(M) is anomalous in the reduce phase space path-integral quantization of general relativity.

\subsection{The path-integral measure from the Holst action}

We apply the general procedure to the case of Holst action, whose canonical framework is studied in \cite{Barros}. The total Hamiltonian is a linear combination of constraints (the physical Hamiltonian on reduced phase space vanishes). the expressions of the first-class constraints $G^{IJ}$, $H_a$, $H$, and second-class constraint $C^{ab}$, $D^{ab}$ is given by (we follow the notation of \cite{Barros})
\be
G_{IJ}&=&\partial_a(\pi-\frac{1}{\g}*\pi)^a_{IJ}+A_{aI}^{\ \ K}(\pi-\frac{1}{\g}*\pi)^a_{JK}-A_{aJ}^{\ \ K}(\pi-\frac{1}{\g}*\pi)^a_{IK}\nonumber\\
H_a&=&\frac{1}{2}(F-\frac{1}{\g}*F)_{ab}^{IJ}[A]\ \pi^b_{IJ}-\frac{1}{2}A^{IJ}_aG_{IJ}\nonumber\\
H&=&\frac{1}{2}(F-\frac{1}{\g}*F)_{ab}^{IJ}[A]\ \pi^a_{IK}\ \pi^b_{JL}\ \eta^{KL}+\l_{ab}(A,\pi)C^{ab}\nonumber\\
C^{ab}&=& \eps^{IJKL}\pi^a_{IJ}\pi^b_{KL}\nonumber\\
D^{ab}&=&*\pi^{c}_{IJ}(\pi^{aIK}D_c\pi^{bJL}+\pi^{bIK}D_c\pi^{aJL})\eta_{KL}
\ee
where $D^{ab}$ is the secondary constraint with $\{H(x),C^{ab}(x')\}=D^{ab}(x)\delta(x,x')$, $\l_{ab}(A,\pi)$ is the Lagrangian multiplier determined by the consistency of $D^{ab}$. The constraint algebra is given by
\be
\left\{G_{IJ}(\L^{IJ}),G_{KL}(\O^{KL})\right\}&=&4G_{IJ}(\L^{IK}\O^{JL}\eta_{KL})\nonumber\\
\left\{G_{IJ}(\L^{IJ}),H_a(N^a)\right\}&=&G_{IJ}(-\cl_{\vec{N}}\L^{IJ})\nonumber\\
\left\{G_{IJ}(\L^{IJ}),H(N)\right\}&=&0\nonumber\\
\left\{G_{IJ}(\L^{IJ}),C^{ab}(c_{ab})\right\}&=&0\nonumber\\
\left\{G_{IJ}(\L^{IJ}),D^{ab}(d_{ab})\right\}&=&0\nonumber\\
\left\{H_a(N^a),H_b(M^b)\right\}&=&H_a(-\cl_{\vec{M}}N^a)\nonumber\\
\left\{H(N),H_a(N^a)\right\}&=&H(-\cl_{\vec{N}}N)\ =\ -H(N^a\partial_aN-N\partial_aN^a)\nonumber\\
\left\{C^{ab}(c_{ab}),H_c(N^c)\right\}&=&C^{ab}(-\cl_{\vec{N}}c_{ab})\nonumber\\
\left\{D^{ab}(d_{ab}),H_c(N^c)\right\}&=&D^{ab}(-\cl_{\vec{N}}d_{ab})\nonumber\\
\left\{H(N),H(M)\right\}&=&H_a([\det q]q^{ab}[N\partial_bM-M\partial_bN])+C^{ab}(\cdots)+D^{ab}(\cdots)\nonumber\\
\left\{H(N),C^{ab}(c_{ab})\right\}&=&D^{ab}(Nc_{ab})+C^{ab}(\cdots)\nonumber\\
\left\{H(N),D^{ab}(d_{ab})\right\}&=&D^{ab}(\cdots)+C^{ab}(\cdots)\nonumber\\
\left\{C^{ab}(c_{ab}),D^{ab}(d_{ab})\right\}&=&C^{ab}(\cdots)+4\left[\det q\right]^2q^{ab}q^{cd}(c_{ac}d_{bd}-c_{ab}d_{cd})
\ee
where we have skipped some unimportant structure functions related to the second-class constraints. Note that the smearing functions $N$ for the Hamiltonian constraint $H$ and $c_{ab}$ for $C^{ab}$ have density weight $-1$, and the smearing function $d_{ab}$ for $D^{ab}$ has density weight $-2$, they are all assumed to be independent of phase space variables. 

In contrast to the ADM formalism, the canonical formulation of Holst action results in a non-regular constrained system (the rank of Dirac matrix is not a constant). There are five disjoint sectors of solutions for the second-class constraint equation $C^{ab}\approx0$, i.e. there exists a non-degenerated triad field $e_a^i$ and an additional 1-form field $e^0_a$ such that 
\begin{eqnarray}
\nonumber
(I \pm) && \pi^a_{IJ}= \pm \eps^{abc} e_b^I e_c^J \nonumber\\
(II \pm) && \pi^a_{IJ} = \pm \frac{1}{2} \eps^{abc} e_b^K e_c^L\eps_{IJKL}\nonumber\\
(\text{Degenerated}) && \pi^a_{IJ} = 0
\end{eqnarray}
where the degenerated sector results in a degenerated Dirac matrix while other sectors lead to non-degenerated Dirac matrix. In order to proceed the path-integral quantization in section \ref{reduce}, we have to exclude the degenerated sector. For simplicity we only consider a single sector, say, II+. The analysis for all other non-degenerated sectors is essentially the same.    

The path-integral partition function can be written down immediately
\be
Z&=&\int\prod_{x\in M}\left[\rmd A_a^{IJ}(x)\ \rmd\pi_{IJ}^a(x)\ \delta\left(C^{ab}(x)\right)\ \delta\left(D^{ab}(x)\right) \sqrt{\big|\det\Delta_D(x)\big|}\ \right]\ \prod_{x\in M}\left[\delta\left(G^{IJ}(x)\right)\ \delta\Big(H_a(x)\Big)\ \delta\Big(H(x)\Big)\right]\nonumber\\
&&\times\prod_{x\in M}\left[\Big|\det\Delta_{FP}(x)\Big|\prod_{\a=1}^9 \delta\Big(\chi^\a\left[A_a^{IJ}(x),\pi^a_{IJ}(x)\right]\Big)\right]\ \exp i\int\rmd t\int\rmd^3x\left[ \frac{1}{2}\pi_{IJ}^a\frac{\partial}{\partial t}\left(A_a^{IJ}-\frac{1}{\gamma}*A_a^{IJ}\right)\right]\nonumber\\
&=&\int\prod_{x\in M}\left[\rmd A_a^{IJ}(x)\ \rmd\pi_{IJ}^a(x)\ \delta\left(C^{ab}(x)\right)\ \delta\left(D^{ab}(x)\right) \sqrt{\big|\det\Delta_D(x)\big|}\ \right]\ \prod_{x\in M}\left[\rmd A_t^{IJ}(x)\ \rmd N^a(x)\ \rmd N(x)\right]\nonumber\\
&&\times\prod_{x\in M}\left[\Big|\det\Delta_{FP}(x)\Big|\prod_{\a=1}^9 \delta\Big(\chi^\a\left[A_a^{IJ}(x),\pi^a_{IJ}(x)\right]\Big)\right]\nonumber\\
&&\times\exp i\int\rmd t\int\rmd^3x\left[ \frac{1}{2}\pi_{IJ}^a\frac{\partial}{\partial t}\left(A_a^{IJ}-\frac{1}{\gamma}*A_a^{IJ}\right)+\frac{1}{2}\left(A_t^{IJ}+N^aA_a^{IJ}\right)G_{IJ}+N^aH_a+NH\right]\label{pi}
\ee
where the first factor in square bracket is the Liouville measure on the constraint surface $(\G_\Psi,\O_\Psi)$ defined by $C^{ab}=D^{ab}=0$. We denote by $D\mu$ the total path-integral measure:
\be
D\mu:=\prod_{x\in M}\left[\rmd A_a^{IJ}(x)\ \rmd\pi_{IJ}^a(x)\ \delta\left(C^{ab}(x)\right)\ \delta\left(D^{ab}(x)\right) \sqrt{\big|\det\Delta_D(x)\big|}\ \right]\ \prod_{x\in M}\left[\rmd A_t^{IJ}(x)\ \rmd N^a(x)\ \rmd N(x)\right]
\ee
$\Delta_D$ is the determinant of the Dirac matrix for the second-class constraint
\be
\begin{pmatrix}
      &\{C^{ab}(x),C^{cd}(x')\}&,\  \{C^{ab}(x),D^{cd}(x')\}\ \  \\
      &\{D^{ab}(x),C^{cd}(x')\}&,\  \{D^{ab}(x),D^{cd}(x')\}\ \ 
\end{pmatrix}
=
\begin{pmatrix}
      &0&,\  \{C^{ab}(x),D^{cd}(x')\}\ \  \\
      &\{D^{ab}(x),C^{cd}(x')\}&,\  \{D^{ab}(x),D^{cd}(x')\}\ \ 
\end{pmatrix}
\ee
Therefore $|\det\Delta_D|=[\det G]^2$ where $G$ is the matrix determined by
\be
G^{ab,cd}(x,x')=\{C^{ab}(x),D^{cd}(x')\}\approx \left[\det q\right]^{2}\left[2q^{ab}q^{cd}-q^{ac}q^{bd}-q^{cb}q^{ad}\right]\ \delta^3(x,x')\label{G}
\ee
its determinant is $\det G=[\det q]^8$ up to overall constant.

As we did before, the first-class constraint algebra on the phase space $(\G_\Psi,\O_\Psi)$ is obtained by computing the Dirac bracket with respect to the second-class constraint
\be
\left\{G_{IJ}(\L^{IJ}),G_{KL}(\O^{KL})\right\}_{\Psi}&\peq&G_{IJ}(4\L^{IK}\O^{JL}\eta_{KL})\nonumber\\
\left\{G_{IJ}(\L^{IJ}),H_a(N^a)\right\}_{\Psi}&\peq&G_{IJ}(-\cl_{\vec{N}}\L^{IJ})\nonumber\\
\left\{G_{IJ}(\L^{IJ}),H(N)\right\}_{\Psi}&\peq&0\nonumber\\
\left\{H_a(N^a),H_b(M^b)\right\}_\Psi&\peq&H_a(-\cl_{\vec{M}}N^a)\nonumber\\
\left\{H(N),H_a(N^a)\right\}_{\Psi}&\peq&H(-\cl_{\vec{N}}N)\ =\ -H(N^a\partial_aN-N\partial_aN^a)\nonumber\\
\left\{H(N),H(M)\right\}_\Psi&\peq&H_a([\det q]q^{ab}[N\partial_bM-M\partial_bN])
\ee
By using this constraint algebra, we are going to derive the local symmetries of the Holst action generated by the first-class constraints, and to check the invariance of total measure $D\mu$ under these local symmetries. So we will see if these classical local symmetries is implemented without anomaly in the reduced phase path-integral quantization. First of all we consider the infinitesimal $SO(\eta)$ gauge transformation $T^\eps_{G}$ generated by $G_{IJ}(\eps^{IJ})$, where $\eps^{IJ}=\eps^{IJ}(t,\vec{x})$ is a spacetime $so(\eta)$-function
\be
f\mapsto T^{\eps}_{G}f:=f+\left\{f,G_{IJ}(\eps^{IJ})\right\}_\Psi\label{Gtrans}
\ee
which results in the transformation of Holst action:
\be
&&S\left[A_a^{IJ},\pi^a_{IJ},A_t^{IJ},N^a,N\right]=\int\rmd t\int\rmd^3x\left[ \frac{1}{2}\pi_{IJ}^a\frac{\partial}{\partial t}\left(A_a^{IJ}-\frac{1}{\gamma}*A_a^{IJ}\right)+\frac{1}{2}\left(A_t^{IJ}+N^aA_a^{IJ}\right)G_{IJ}+N^aH_a+NH\right]\nonumber\\
&\mapsto& S\left[T^\eps_{G}{A}_a^{IJ},T^\eps_{G}{\pi}^a_{IJ},T^\eps_{G}{A}_t^{IJ},T^\eps_{G}{N}^a,T^\eps_{G}{N}\right]\nonumber\\&&=\int\rmd t\int\rmd^3x\Bigg[\frac{1}{2}\pi_{IJ}^a\frac{\partial}{\partial t}\left(A_a^{IJ}-\frac{1}{\gamma}*A_a^{IJ}\right)+\frac{1}{2}\Bigg(T^\eps_{G}\left(A_t^{IJ}+N^aA_a^{IJ}\right)G_{IJ}+4\left(A_t^{IK}+N^aA_a^{IK}\right)\eps^{JL}\eta_{KL}G_{IJ}\Bigg)\nonumber\\
&&+\ \Bigg(T^\eps_{G}N^aH_a+\cl_{\vec{N}}\eps^{IJ}G_{IJ}\Bigg)+T^\eps_{G}NH\Bigg]
\ee  
If we suppose the Lagrangian multipliers obey the following transformation,
\be
{A}_t^{IJ}&\mapsto&T^\eps_{G}{A}_t^{IJ}:=A_t^{IJ}-4A_t^{[I|K|}\eps^{J]L}\eta_{KL}-2\partial_t\eps^{IJ}\nonumber\\
N^a&\mapsto&T^\eps_{G}N^a:=N^a\nonumber\\
N&\mapsto&T^\eps_{G}N:=N\label{Gtrans1}
\ee
the Holst action stays unchanged: $S\left[T^\eps_{G}{A}_a^{IJ},T^\eps_{G}{\pi}^a_{IJ},T^\eps_{G}{A}_t^{IJ},T^\eps_{G}{N}^a,T^\eps_{G}{N}\right]=S\left[A_a^{IJ},\pi^a_{IJ},A_t^{IJ},N^a,N\right]$ by using the consistency condition $\partial_t{G}^{IJ}=0$ and the gauge transformation of $A_a^{IJ}$
\be
{A}_a^{IJ}&\mapsto&T^\eps_{G}{A}_a^{IJ}=A_a^{IJ}+\left\{A_a^{IJ}(x),G_{IJ}(\eps^{IJ})\right\}_\Psi=A_a^{IJ}-4A_a^{[I|K|}\eps^{J]L}\eta_{KL}-2\partial_a\eps^{IJ}.
\ee
Thus we obtain the SO($\eta$) local symmetry of the Holst action generated by Gauss constraint. It is easy to see that we have recovered the SO($\eta$) local gauge transformation of the spacetime connection field, i.e.
\be
{A}_\a^{IJ}&\mapsto&T^\eps_{G}{A}_\a^{IJ}=A_\a^{IJ}-4A_\a^{[I|K|}\eps^{J]L}\eta_{KL}-2\partial_\a\eps^{IJ}.
\ee 
Since the Jacobian matrix from Eq.(\ref{Gtrans1}) is independent of the phase space variables, the total path-integral measure $D\mu$ is invariant (up to an overall constant) under $T^\eps_G$.

Next we analyze the infinitesimal gauge transformation $T^\eps_{Diff}$ generated by the spatial diffeomorphism constraint $H_{a}(\eps^{a})$, $\eps^a=\eps^a(t,\vec{x})$. For any phase space function, the infinitesimal gauge transformation generated by $H_a(\eps^a)$ reads 
\be
f\mapsto T^{\eps}_{Diff}f:=f+\left\{f,H_{a}(\eps^{a})\right\}_\Psi\label{Difftrans}
\ee
which results in the transformation of Holst action:
\be
&&S\left[A_a^{IJ},\pi^a_{IJ},A_t^{IJ},N^a,N\right]=\int\rmd t\int\rmd^3x\left[ \frac{1}{2}\pi_{IJ}^a\frac{\partial}{\partial t}\left(A_a^{IJ}-\frac{1}{\gamma}*A_a^{IJ}\right)+\frac{1}{2}\left(A_t^{IJ}+N^aA_a^{IJ}\right)G_{IJ}+N^aH_a+NH\right]\nonumber\\
&\mapsto& S\left[T^{\eps}_{Diff}{A}_a^{IJ},T^{\eps}_{Diff}{\pi}^a_{IJ},T^{\eps}_{Diff}{A}_t^{IJ},T^{\eps}_{Diff}{N}^a,T^{\eps}_{Diff}{N}\right]\nonumber\\
&&=\int\rmd t\int\rmd^3x\Bigg[\frac{1}{2}\pi_{IJ}^a\frac{\partial}{\partial t}\left(A_a^{IJ}-\frac{1}{\gamma}*A_a^{IJ}\right)+\frac{1}{2}\left[T^{\eps}_{Diff}A_t^{IJ}-\cl_{\vec{\eps}}A_t^{IJ}+\left(T^{\eps}_{Diff}N^a\right)A_a^{IJ}+N^a\cl_{\vec{\eps}}A_a^{IJ}-\cl_{\vec{\eps}}\left(N^aA_a^{IJ}\right)\right]G_{IJ}\nonumber\\
&&+\ \Big[T^{\eps}_{Diff}N^a-\cl_{\vec{\eps}}N^a\Big]H_a+\Big[T^{\eps}_{Diff}N-\cl_{\vec{\eps}}N\Big]H\ \Bigg]
\ee  
Obviously, in order that $\delta_{Diff,\eps}S=0$, the infinitesimal transformation of Lagrangian multipliers takes the following form:
\be
{A}_t^{IJ}&\mapsto&T^\eps_{Diff}{A}_t^{IJ}:=A_t^{IJ}+\cl_{\vec{\eps}}A_t^{IJ}-A_a^{IJ}\partial_t\eps^a=A_t^{IJ}+\eps^a\partial_aA_t^{IJ}-A_a^{IJ}\partial_t\eps^a\nonumber\\
N^a&\mapsto&T^\eps_{Diff}N^a:=N^a+\cl_{\vec{\eps}}N^a+\partial_t\eps^a=N^a+\eps^b\partial_bN^a-N^b\partial_b\eps^a+\partial_t\eps^a\nonumber\\
N&\mapsto&T^\eps_{Diff}N:=N+\cl_{\vec{\eps}}N=N+\eps^a\partial_aN-N\partial_a\eps^a\label{Difftrans1}
\ee
note that here the lapse $N$ has density-weight -1. Thus we obtain the local symmetry of the Holst action generated by spatial diffeomorphism constraint. Again since the Jacobian matrix from Eq.(\ref{Difftrans1}) is independent of phase space variables, the total path-integral measure $D\mu$ is invariant (up to an overall constant) under $T^\eps_{Diff}$.

The last task is to consider the infinitesimal gauge transformation $T^\eps_{H}$ generated by the Hamiltonian constraint $H(\eps)$ with $\eps=\eps(t,\vec{x})$. For any phase space function, the infinitesimal gauge transformation generated by $H(\eps)$ reads 
\be
f\mapsto T^{\eps}_{H}f:=f+\left\{f,H(\eps)\right\}_\Psi\label{Htrans}
\ee
which results in the transformation of the Holst action:
\be
&&S\left[A_a^{IJ},\pi^a_{IJ},A_t^{IJ},N^a,N\right]=\int\rmd t\int\rmd^3x\left[ \frac{1}{2}\pi_{IJ}^a\frac{\partial}{\partial t}\left(A_a^{IJ}-\frac{1}{\gamma}*A_a^{IJ}\right)+\frac{1}{2}\left(A_t^{IJ}+N^aA_a^{IJ}\right)G_{IJ}+N^aH_a+NH\right]\nonumber\\
&\mapsto& S\left[T^{\eps}_{H}{A}_a^{IJ},T^{\eps}_{H}{\pi}^a_{IJ},T^{\eps}_{H}{A}_t^{IJ},T^{\eps}_{H}{N}^a,T^{\eps}_{H}{N}\right]\nonumber\\
&&=\int\rmd t\int\rmd^3x\Bigg[\frac{1}{2}\pi_{IJ}^a\frac{\partial}{\partial t}\left(A_a^{IJ}-\frac{1}{\gamma}*A_a^{IJ}\right)+\frac{1}{2}\left[T^{\eps}_{H}A_t^{IJ}+\left(T^{\eps}_{H}N^a\right)A_a^{IJ}+N^a\left\{A_a^{IJ},H(\eps)\right\}_\Psi\right]G_{IJ}\nonumber\\
&&+\ \left(T^{\eps}_{H}N^a\right)H_a+\left(\cl_{\vec{N}}\eps\right) H+\left(T^{\eps}_{H}N\right)H+[\det q]q^{ab}(N\partial_b\eps-\eps\partial_bN)H_a\Big]\Bigg]
\ee  
in order that the action is invariant $\delta_{H,\eps}S=0$, the infinitesimal transformations of the Lagrangian multipliers takes the following form, in order to obtain the local symmetries of the Holst action generated by the Hamiltonian constraint $H$
\be
{A}_t^{IJ}&\mapsto&T^\eps_{H}{A}_t^{IJ}:=A_t^{IJ}+[\det q]q^{ab}\left(N\partial_b\eps-\eps\partial_bN\right)A_a^{IJ}-N^a\left\{A_a^{IJ},H(\eps)\right\}_\Psi\nonumber\\
N^a&\mapsto&T^\eps_{H}N^a:=N^a-[\det q]q^{ab}\left(N\partial_b\eps-\eps\partial_bN\right)\nonumber\\
N&\mapsto&T^\eps_{H}N:=N-\cl_{\vec{N}}\eps+\partial_t\eps=N-\left(N^a\partial_a\eps-\eps\partial_aN^a\right)+\partial_t\eps\label{Htrans1}
\ee
note that here the lapse $N$ has density-weight -1. the Jacobian matrix $\partial \left(T^\eps_{H}{A}_t^{IJ},T^\eps_{H}N^a,T^\eps_{H}N\right)/\partial\left({A}_t^{IJ},N^a,N\right)$ for the transformation Eq.(\ref{Htrans1}) is expressed as
\begin{center}
\begin{tabular}{|c||c|c|c|}
\hline
$\frac{\partial \left(T^\eps_{H}{A}_t^{IJ},T^\eps_{H}N^a,T^\eps_{H}N\right)}{\partial\left({A}_t^{IJ},N^a,N\right)}$ & $T^\eps_{H}{A}_t^{IJ}$ & $T^\eps_{H}N^a$ & $T^\eps_{H}N$ \\
\hline\hline
$A_t^{IJ}$ & $1$ & $0$ & $0$ \\
\hline
$N^a$ &$\left\{A_a^{IJ},H(\eps)\right\}_\Psi$ & $1$ & $-\left(\partial_a\eps-\eps\partial_a\right)$\\
\hline
$N$& $-[\det q]q^{ab}A_a^{IJ}\left(\eps\partial_b-\partial_b\eps\right)$ & $[\det q]q^{ab}\left(\eps\partial_b-\partial_b\eps\right)$ & $1$\\
\hline
\end{tabular}
\vspace{0.5cm}\end{center}
which can be written as an identity matrix with the addition of a perturbation, i.e.
\be
\frac{\partial \left(T^\eps_{H}{A}_t^{IJ},T^\eps_{H}N^a,T^\eps_{H}N\right)}{\partial\left({A}_t^{IJ},N^a,N\right)}=I+\begin{pmatrix}
    0  &  0  &  0\\
    \ \left\{A_a^{IJ},H(\eps)\right\}_\Psi  & 0   &\ -\left(\partial_a\eps-\eps\partial_a\right) \\
    \ -[\det q]q^{ab}A_a^{IJ}\left(\eps\partial_b-\partial_b\eps\right) &\ [\det q]q^{ab}\left(\eps\partial_b-\partial_b\eps\right)   & 0
\end{pmatrix}\equiv I+\ce
\ee
The perturbation $\ce$ is again traceless, so the determinant of the Jacobian matrix equals $1+\mathrm{Tr}\ce=1$, which means that the local symmetries of the Holst action generated by $H$, Eqs.(\ref{Htrans}) and (\ref{Htrans1}), leave the total path-integral measure $D\mu$ invariant.

The result of this section is summarized in the following theorem:
\begin{Theorem}\label{invmeasure}
(1) The infinitesimal transformations induced by $SO(\eta)$ Gauss constraint $G^{IJ}$
\be
f\left[A_a^{IJ},\pi^a_{IJ}\right]&\mapsto& T^{\eps}_{G}f\left[A_a^{IJ},\pi^a_{IJ}\right]:=f\left[A_a^{IJ},\pi^a_{IJ}\right]+\left\{f\left[A_a^{IJ},\pi^a_{IJ}\right],G_{IJ}(\eps^{IJ})\right\}_\Psi\nonumber\\
{A}_t^{IJ}&\mapsto&T^\eps_{G}{A}_t^{IJ}:=A_t^{IJ}-4A_t^{[I|K|}\eps^{J]L}\eta_{KL}+2\partial_t\eps^{IJ}\nonumber\\
N^a&\mapsto&T^\eps_{G}N^a:=N^a\nonumber\\
N&\mapsto&T^\eps_{G}N:=N
\ee
form a group of SO($\eta$) local symmetries of the Holst action and leave the path-integral measure $D\mu$ invariant (up to an overall constant). So the classical SO($\eta$) local symmetries of the Holst action is implemented without anomaly in reduced phase space path-integral quantization.

(2) There exist infinitesimal transformations $T^{\eps^\mu}$ ($\eps^\mu=\eps^\mu(t,\vec{x})$) induced by the spatial diffeomorphism constraint $H_a$ and the Hamiltonian constraint $H$
\be
f\left[A_a^{IJ},\pi^a_{IJ}\right]&\mapsto& T^{\eps^\mu}f\left[A_a^{IJ},\pi^a_{IJ}\right]:=f\left[A_a^{IJ},\pi^a_{IJ}\right]+\left\{f\left[A_a^{IJ},\pi^a_{IJ}\right],H(\eps)+H_a(\eps^{a})\right\}_\Psi\nonumber\\
{A}_t^{IJ}&\mapsto&T^{\eps^\mu}{A}_t^{IJ}:=A_t^{IJ}+[\det q]q^{ab}\left(N\partial_b\eps-\eps\partial_bN\right)A_a^{IJ}-N^a\left\{A_a^{IJ},H(\eps)\right\}_\Psi+\cl_{\vec{\eps}}A_t^{IJ}-A_a^{IJ}\partial_t\eps^a\nonumber\\
N^a&\mapsto&T^{\eps^\mu}N^a:=N^a-[\det q]q^{ab}\left(N\partial_b\eps-\eps\partial_bN\right)+\cl_{\vec{\eps}}N^a+\partial_t\eps^a\nonumber\\
N&\mapsto&T^{\eps^\mu}N:=N+\cl_{\vec{\eps}}N-\cl_{\vec{N}}\eps+\partial_t\eps
\ee
The enveloping algebra generated by $\{T^{\eps^\mu}\}_{\eps^\mu}$ is the Bergmann-Komar group BK(M) represented on the space of connections and tetrads. Each element of BK(M) is an infinitesimal local symmetry of the Holst action $\delta_{\eps^\mu}S=0$. And the total path-integral measure $D\mu$ is invariant (up to an overall constant) under these transformations. Thus the Bargmann-Komar group BK(M) as a collection of classical local symmetries is implemented without anomaly in the reduced phase space path-integral quantization.
\end{Theorem}

\section{Path-integral and refined algebraic quantization}\label{PIRAQ}

Since the dynamics of GR is completely determined by the constraints, the criterion for the correctness of a path-integral is that it should solve all the constraints of GR quantum mechanically. In the following analysis, we will discuss this point in the framework of refined algebraic quantization (RAQ).

\subsection{Refined algebraic quantization}

Consider a general constrained dynamical system with a number of the first-class constraints $C_I$, we denote the full phase space by $(\cm,\o)$ and the canonical coordinate on it by $(p_a,q^a)$. In the procedure of canonical quantization of the constrained system, it is practically simpler to first perform the quantization on the phase space $(\cm,\o)$, when one doesn't have enough knowledges about the reduced phase space variables. Suppose we have done this quantization and obtained a kinematical Hilbert space $\ch_{Kin}$, which is a space of the $L^2$-functions of configuration variables $q^a$. After that, one should represent the constraints as the operators $\hat{C}_I$ on $\ch_{Kin}$, and impose the quantum constraint equation $\hat{C}_I\Psi=0$. In general the solution of the constraint equations doesn't belong to the kinematical Hilbert space, but will contained by the algebraic dual $\Fd^\star_{Kin}$ of a dense domain $\Fd_{Kin}\subset\ch_{Kin}$, which is supposed to be invariant under all the $\hat{C}_I$ and $\hat{C}_I^\dagger$. So what we are looking for is the state $\Psi\in\Fd^\star$ such that:
\be
\Psi\lt[\hat{C}^\dagger_If\rt]:=\hat{C}'_I\Psi\lt[f\rt]=0,\ \ \ \ \ \ \forall f\in\Fd
\ee
The space of solutions is denoted by $\Fd^\star_{Phys}$. The physical Hilbert space will be a subspace of $\Fd^\star_{Phys}$. And $\Fd^\star_{Phys}$ will be the algebraic dual of a dense domain $\Fd_{Phys}\in\ch_{Phys}$, which is the invariant domain of the algebra of operators corresponding to the Dirac observables. Hence we obtain a Gel'fand triple:
\be
\Fd_{Phys}\hookrightarrow\ch_{Phys}\hookrightarrow \Fd^\star_{Phys}
\ee

A systematic construction of the physical Hilbert space is available if we have an anti-linear rigging map:
\be
\eta:\Fd_{Kin}\to\Fd_{Phys}^\star;\ f\mapsto\eta(f)
\ee
such that 
(1.) $\eta(f')[f]$ is a positive semi-definite sesquilinear form on $\Fd_{Kin}$. 
(2.) For all the Dirac observables $\hat{O}$ on the kinematical Hilbert space, we have $\hat{O}'\eta(f)=\eta(\hat{O}f)$. 
If such a rigging map $\eta$ exists, we define the physical inner product by 
\be
\lag\eta(f)|\eta(f')\rag_{Phys}:=\eta(f')[f],\ \forall f,f'\in\Fd_{Kin}.
\ee
Then a null space $\Fn\subset\Fd_{Phys}^\star$ is defined by $\lt\{\eta(f)\in\Fd_{Phys}^\star\ \big|\ ||\eta(f)||_{Phys}=0\ \rt\}$. Therefore 
\be
\Fd_{Phys}:=\eta\lt(\Fd_{Kin}\rt)/\Fn
\ee
And the physical Hilbert space $\ch_{Phys}$ is defined by the completion of $\Fd_{Phys}$ with respect to the physical inner product.

The above general procedure is called the Refined Algebraic Quantization (RAQ) \cite{RAQ}, which follows the Dirac quantization procedure of a first-class constrained system, and independent from the reduced phase space quantization described in section \ref{reduce}.

\subsection{The path-integral as a rigging map}\label{PIARM}

The discussion in section \ref{reduce} suggests a anti-linear map $\eta^{\t_f,\t_i}_\o$ from $\Fd_{Kin}$ to $\Fd^\star_{Kin}$, by a choice of the dense domain $\Fd_{Kin}\subset\ch_{Kin}$, two multi-finger clock values $\t_f, \t_i$, and a reference vector $\o\in\Fd_{Kin}$:
\be
&&\eta^{\t_f,\t_i}_\o: \Fd_{Kin}\to\Fd_{Kin}^\star;\ f\mapsto\eta^{\t_f,\t_i}_\o(f)\nonumber\\
&&\eta^{\t_f,\t_i}_\o(f)[f']:=\frac{\int \cd p_a\cd q^a\cd P_I\cd T^I \prod_{t,I}\left[\delta\Big(P_I+h_I\Big)\ \delta\Big(T^I-c^I\Big)\right]\ e^{ i\int_{t_i}^{t_f}\rmd t\left[\sum_ap_a(t)\dot{q}^a(t)+\sum_IP_I(t)\dot{T}^I(t) \right]}\ \overline{f(q^a_f,T^I_f)}f'(q^a_f,T^I_f)}{\int \cd p_a\cd q^a\cd P_I\cd T^I \prod_{t,I}\left[\delta\Big(P_I+h_I\Big)\ \delta\Big(T^I-c^I\Big)\right]\ e^{ i\int_{t_i}^{t_f}\rmd t\left[\sum_ap_a(t)\dot{q}^a(t)+\sum_IP_I(t)\dot{T}^I(t) \right]}\ \overline{\o(q^a_f,T^I_f)} {\o}(q^a_i,T^I_i)}
\ee
for all $f,f'\in\Fd_{Kin}$. As it is mentioned in section \ref{reduce}, the path-integral definition of $\eta^{\t_f,\t_i}_\o(f)[f']$ doesn't depend on the choice of path $c$ in the $\ct$-space. However, for general kinematical state $f,f'$, $\eta^{\t_f,\t_i}_\o(f)[f']$ depends on the choice of the initial and final points of $c$, i.e. it depends on the value of $\t_i=c(t_i)$ and $\t_f=c(t_f)$. The reason is the following: given a specific $\t_0$ and a kinematical state $f\in\Fd_{Kin}$, we restrict the function $f(q^a,T^I)$ on the surface defined by $T^I-\t^I_0=0$ and keep in mind that $q^a=Q^a(0)\equiv Q^a$. Then we obtain a wave-packet $f(Q^a,\t_0)$ at the multi-finger time $\t_0$. This wave-packet belongs to the Hilbert space $\ch$ from reduced phase space quantization, and serves as the initial wave-packet for multi-finger time evolution. That is, we obtain a wave function $F_{\t_0}(Q^a,\t)$ by solving the Sch\"odinger equation for each $I$:
\be
i\frac{\partial}{\partial\t^I}F_{\t_0}(Q^a,\t)=\hat{H}_I(\t)F_{\t_0}(Q^a,\t)
\ee
with the initial condition:
\be
F_{\t_0}(Q^a,\t_0)=f(Q^a,\t_0)
\ee
The multi-finger history of the solution $F_{\t_0}(Q^a,\t)$ corresponds to a Heisenberg state $|F_{\t_0}\rag\in\ch$ (we use the lower case letter to denote the kinematical state in $\ch_{Kin}$, while the corresponding state in $\ch$ is denoted by the corresponding capital letter). Note that the above map from kinematical states to the Heisenberg states in $\ch$ is not injective for a given $\t_0$, since different kinematical functions can have the same restriction on the surface $T^I-\t^I_0=0$. As a result, by reversing the calculation in section \ref{reduce}, we see that
\be
\eta^{\t_f,\t_i}_\o(f)[f']=\frac{\lag F_{\t_f}|F'_{\t_i}\rag}{\lag \O_{\t_f}|\O_{\t_i}\rag}
\ee
which shows the dependence on $\t_f$ and $\t_i$. However, for the purpose of the following analysis, we remove this dependence by integral over all possible $\t_f$ and $\t_i$, and define a new anti-linear map $\eta_\o$, which is the candidate for a formal rigging map:
\be
&&\eta_\o: \Fd_{Kin}\to\Fd_{Kin}^\star;\ f\mapsto\eta_\o(f)\nonumber\\
&&\eta_\o(f)[f']:=\frac{\int\prod_{t=[t_i,t_f]}\rmd c(t)\ \lag F_{\t_f}|F'_{\t_i}\rag}{\int\prod_{t=[t_i,t_f]}\rmd c(t)\ \lag \O_{\t_f}|\O_{\t_i}\rag}=\frac{\int\rmd\t_f\rmd\t_i\ \lag F_{\t_f}|F'_{\t_i}\rag}{\int\rmd\t_f\rmd\t_i\ \lag \O_{\t_f}|\O_{\t_i}\rag}\nonumber\\
&=&\frac{\int \cd p_a\cd q^a\cd P_I\cd T^I \prod_{t,I}\delta\Big(P_I+h_I\Big)\ e^{ i\int_{t_i}^{t_f}\rmd t\left[\sum_ap_a(t)\dot{q}^a(t)+\sum_IP_I(t)\dot{T}^I(t) \right]}\ \overline{f(q^a_f,T^I_f)}f'(q^a_f,T^I_f)}{\int \cd p_a\cd q^a\cd P_I\cd T^I \prod_{t,I}\delta\Big(P_I+h_I\Big)\ e^{ i\int_{t_i}^{t_f}\rmd t\left[\sum_ap_a(t)\dot{q}^a(t)+\sum_IP_I(t)\dot{T}^I(t) \right]}\ \overline{\o(q^a_f,T^I_f)} {\o}(q^a_i,T^I_i)}
\ee
where we use the fact that $\lag F_{\t_f}|F'_{\t_i}\rag$ is independent of the path $c(t)$ except its initial and final points, so the integrals $\int\prod_{t=[t_i,t_f]}\rmd c(t)$ are canceled from the denominator and numerator except $\int\rmd\t_f\rmd\t_i$. Suppose $\eta_\o(f)[f']$ is finite for all $f,f'\in\Fd_{Kin}$, $\eta_{\o}(f)$ then is an element in $\Fd^\star_{Kin}$. If $\eta_\o$ is a rigging map, the image of $\eta_\o$ should be in $\Fd_{Phys}^\star$, we have to first check if $\eta_\o(f)$ solves all the quantum constraint equations.

We suppose all the constraints $C_I$ have been Abelianized as $C'_I=P_I+h_I(p_a,q^a,T^I)$. The corresponding operators is defined by
\be
\hat{C}'_If(q^a,T^I):=\lt[-i\frac{\partial}{\partial T^I}+h_I\lt(-i\frac{\partial}{\partial q^a},q^a,T^I\rt)\rt]\ f(q^a,T^I)
\ee
Then we compute $\eta_\o(f)[\hat{C}'_If']$ for any two kinematical states $f,f'\in\Fd_{Kin}$
\be
&&\int \cd p_a(t)\cd q^a(t)\cd P_I(t)\cd T^I(t)\ \prod_{t,I}\delta\Big(P_I+h_I\Big)\ e^{ i\int_{t_i}^{t_f}\rmd t\left[\sum_ap_a(t)\dot{q}^a(t)+\sum_IP_I(t)\dot{T}^I(t) \right]}\ \overline{f(q^a_f,T^I_f)}\ \hat{C}'_If'(q^a_i,T^I_i)\nonumber\\
&=&\int \cd p_a(t)\cd q^a(t)\cd T^I(t)\ e^{ i\int_{t_i}^{t_f}\rmd t\left[\sum_ap_a(t)\dot{q}^a(t)+\sum_Ih_I(t)\dot{T}^I(t) \right]}\ \overline{f(q^a_f,T^I_f)}\ \lt[-i\frac{\partial}{\partial T^I_i}+h_I\lt(-i\frac{\partial}{\partial q_i^a},q_i^a,T_i^I\rt)\rt]\ f'(q^a_i,T^I_i)
\ee
We now consider the collection of $T^I$ as the multi-finger time parameters, then the constraint equations
\be
\hat{C}'_IF(q^a,T^I):=\lt[-i\frac{\partial}{\partial T^I}+h_I\lt(-i\frac{\partial}{\partial q^a},q^a,T^I\rt)\rt]\ F(q^a,T^I)=0
\ee
can be consider as the multi-finger time evolution equations. Similar to the case with $\t$-parameter, given a kinematical state $f$, a specific $T_0$ and the initial condition $F(q^a,T_0^I)=f(q^a,T_0^I)$, a solution $F_{T_0}(q^a,T^I)$ can be determined in principle and corresponds to a state (a Sch\"odinger state $|F_{T_0}(T)\rag$ or a Heisenberg state $|F_{T_0}\rag$ ) in $\ch$. And an unitary propagator $U(T_f,T_i)$ is defined on $\ch$ in the same way as it was in section \ref{reduce}. Therefore, we have
\be
&&\int \cd p_a(t)\cd q^a(t)\cd P_I(t)\cd T^I(t)\ \prod_{t,I}\delta\Big(P_I+h_I\Big)\ e^{ i\int_{t_i}^{t_f}\rmd t\left[\sum_ap_a(t)\dot{q}^a(t)+\sum_IP_I(t)\dot{T}^I(t) \right]}\ \overline{f(q^a_f,T^I_f)}\ \hat{C}'_If'(q^a_i,T^I_i)\nonumber\\
&=&\int\cd T^I(t)\lt\lag F_{T_f}(T_f)\Bigg|U(T_f,T_i)\lt[-i\frac{\rmd}{\rmd T_i^I}+\hat{h}(T_i)\rt]\Bigg|F'_{T_i}(T_i)\rt\rag\nonumber\\
&=&\int\cd T^I(t)\Bigg(i\frac{\rmd}{\rmd T_i^I}\Big\lag F_{T_f}(T_f)\Big|U(T_f,T_i)\Bigg)\ \Big|F'_{T_i}(T_i)\Big\rag+\Big\lag F_{T_f}(T_f)\Big|U(T_f,T_i)\ \hat{h}(T_i)\Big|F'_{T_i}(T_i)\Big\rag\nonumber\\
&=&0
\ee
by some certain boundary conditions of $f(q^a_i,T^I_i)$ and $f'(q^a_i,T^I_i)$, note that $\cd T^I(t):=\prod_{t\in[t_i,t_f]}\rmd T^I(t)$. The above manipulation shows that
\be
\eta_\o(f)[\hat{C}'_If']=0.
\ee
So the map $\eta_\o$ is from $\Fd_{Kin}$ to the space of solutions $\Fd_{Phys}^\star$ of the quantum constraint equations.

Moreover, it is clear that $\eta_\o(f)[f']$ is a positive semi-definite sesquilinear form on $\Fd_{Kin}$ because by definition
\be
\eta_\o(f)[f']=\frac{\int\rmd\t_1\rmd\t_2\ \lag F_{\t_1}|F_{\t_2}\rag}{\int\rmd\t_1\rmd\t_2\ \lag \O_{\t_1}|\O_{\t_2}\rag}=\frac{\int\rmd Q^a \overline{\lt[\int\rmd\t_1\ F_{\t_1}(Q^a)\rt]}\ \lt[\int\rmd\t_2\ F_{\t_2}(Q^a)\rt]}{\int\rmd Q^a \overline{\lt[\int\rmd\t_1\ \O_{\t_1}(Q^a)\rt]}\ \lt[\int\rmd\t_2\ \O_{\t_2}(Q^a)\rt]}
\ee
Therefore we define formally the physical inner product via the map $\eta_\o$
\be
\lag\eta_\o(f')|\eta_\o(f)\rag_{Phys}&:=&\eta_\o(f)[f']\nonumber\\
&=&\frac{\int \cd p_a\cd q^a\cd P_I\cd T^I \prod_{t,I}\delta\Big(P_I+h_I\Big)\ e^{ i\int_{t_i}^{t_f}\rmd t\left[\sum_ap_a(t)\dot{q}^a(t)+\sum_IP_I(t)\dot{T}^I(t) \right]}\ \overline{f(q^a_f,T^I_f)}f'(q^a_i,T^I_i)}{\int \cd p_a\cd q^a\cd P_I\cd T^I \prod_{t,I}\delta\Big(P_I+h_I\Big)\ e^{ i\int_{t_i}^{t_f}\rmd t\left[\sum_ap_a(t)\dot{q}^a(t)+\sum_IP_I(t)\dot{T}^I(t) \right]}\ \overline{\o(q^a_f,T^I_f)} {\o}(q^a_i,T^I_i)}\nonumber\\
&\equiv&\frac{1}{Z_\o}{\int \cd p_a\cd q^a\cd P_I\cd T^I \prod_{t,I}\delta\Big(P_I+h_I\Big)\ e^{ i\int_{t_i}^{t_f}\rmd t\left[\sum_ap_a(t)\dot{q}^a(t)+\sum_IP_I(t)\dot{T}^I(t) \right]}\ \overline{f(q^a_f,T^I_f)}f'(q^a_i,T^I_i)}\label{rigging}
\ee
On the other hand, the physical inner product Eq.(\ref{rigging}) can also be equivalently expressed by the group averaging of the Abelianized constraints, i.e. if the Abelianized constraint ${C}'_I$ are represented as commutative self-adjoint operators $\hat{C}'_I$, 
\be
\eta_\o(f)[f']=\frac{\int\prod_{I}\rmd t^I\lt\lag f \lt|\ \exp\lt[i\sum_It^I\hat{C}'_I\rt]\ \rt| f'\rt\rag}{\int\prod_{I}\rmd t^I\lt\lag \o \lt|\ \exp\lt[i\sum_It^I\hat{C}'_I\rt]\ \rt| \o\rt\rag}
\ee 
The formal proof for this equivalence is shown in \cite{muxin}. From this equation, it is clear that for any operator $\hat{O}$ such that its adjoint $\hat{O}^\dagger$ commutating with all the constraints $\hat{C}_I'$
\be
\hat{O}'\eta_\o(f)[f']=\eta_\o(f)[\hat{O}^\dagger f']=\eta_\o(\hat{O}f)[f']
\ee
Therefore $\eta_\o$ satisfies all the requirements, thus is qualified as a rigging map.

The physical Hilbert space $\ch_{Phys}$ is defined by the image of $\eta_{\o}$ modulo a null space of zero-norm states, while the detailed structure of the null space and physical Hilbert space $\ch_{Phys}$ should be studied case by case.

Before we come to the next section, it is remarkable that Eq.(\ref{rigging}) removes all the gauge fixing conditions from the original path-integral formula Eq.(\ref{RPI3}). It will greatly simplify the path-integral of gravity, since the gauge fixings are often hard to implement in the case of quantum gravity.

\subsection{The case of gravity with scalar fields}

We now apply our general consideration to the case of gravity with four massless real scalar fields $T^I$ $(I=0,\cdots,3)$, where the action for the scalar fields reads
\be
S_{KG}[T^I,g_{\mu\nu}]=-\frac{\a}{2}\sum_{I=0}^3\int\rmd^4x\ \sqrt{|\det(g)|}\ g^{\a\b}\partial_\a T^I \partial_\b T^I\label{dustaction}
\ee
where $\a$ is the coupling constraint.

One can perform the Legendre transformation of Eq.(\ref{dustaction}) according to the 3+1 decomposition of the spacetime manifold $M\simeq\mathbb{R}\times\Sig$. The resulting total Hamiltonian of the system is a linear combination of first class constraints $C^{tot}$ and $C^{tot}_a$:
\be
C^{tot}&=&C+C^{KG},\ \ \ \ \ \ \ \ C^{KG}\ =\ \frac{1}{\sqrt{\det q}}\lt[\frac{\a(\det q)}{2}q^{ab}\partial_aT^I\partial_bT^I+\frac{1}{2\a}P_I^2\rt]\nonumber\\
C^{tot}_a&=&C_a+C^{KG}_a,\ \ \ \ \ \ C_a^D\ =\ P_I\partial_aT^I\label{constraint}
\ee
here $C$ and $C_a$ are respectively the standard Hamiltonian constraint and diffeomorphism constraint for gravity. The symplectic structure here is different from the standard case of gravity only by adding the scalar variables $T^I$ and their conjugate momenta $P_I$. 

When the local Jacobian matrix
\be
\frac{\partial \lt(C^{tot},C^{tot}_a\rt)}{\partial\lt(P_0,P_j\rt)}=
\left(\begin{array}{cc}
\frac{P_0}{\a\sqrt{\det q}} & \frac{P_j}{\a\sqrt{\det q}} \\
\partial_aT^0 & \partial_a T^j
\end{array}\right)
\ee
is non-degenerated, the constraint Eqs.(\ref{constraint}) can be locally solved into the equivalent form as Eq.(\ref{solve})
\be
\tilde{C}^{tot}&=&P+h,\ \ \ \ \ \ \ \ h\ =\ h(P^{ab},q^{ab},T^I)\nonumber\\
\tilde{C}^{tot}_j&=&P_j+h_j,\ \ \ \ \ \ h_j\ =\ h_j(P^{ab},q^{ab},T^I).\label{csolve}
\ee
Since $C^{tot}$ is quadratic to $P$, we have to restrict $P$ to take value only in half real-line in order to obtain $\tilde{C}^{tot}$. The constraint Eqs.(\ref{csolve}) form a strongly Abelean constraint algebra.

Following the general rule in section \ref{PIARM}, we can immediately write down the formal rigging map. We first consider the canonical GR in the ADM formalism
\be
&&\eta_\o: \Fd_{Kin}\to\Fd_{Phys}^\star;\ f\mapsto\eta_\o(f),\ \ \eta_\o(f)[f']:=\nonumber\\
&&\frac{\int \cd P^{ab}\cd q_{ab}\cd P_I\cd T^I \prod_{x\in M}\delta\Big(\tilde{C}^{tot}\Big)\ \delta\Big(\tilde{C}^{tot}_j\Big)\ e^{ i\int_{t_i}^{t_f}\rmd t\int_{\Sig_t}\rmd^3x\left[P^{ab}\dot{q}_{ab}+P_I\dot{T}^I\right]}\ \overline{f\lt(q_{ab},T^I\rt)_{t_f}}f'\lt(q_{ab},T^I\rt)_{t_i}}{\int \cd P^{ab}\cd q_{ab}\cd P_I\cd T^I \prod_{x\in M}\delta\Big(\tilde{C}^{tot}\Big)\ \delta\Big(\tilde{C}^{tot}_j\Big)\ e^{ i\int_{t_i}^{t_f}\rmd t\int_{\Sig_t}\rmd^3x\left[P^{ab}\dot{q}_{ab}+P_I\dot{T}^I\right]}\ \overline{\o\lt(q_{ab},T^I\rt)_{t_f}}\o\lt(q_{ab},T^I\rt)_{t_i}} \label{ADMrigging}
\ee
Note that the integral of $P$ is only over a half of the real-line.

The formal rigging map Eq.(\ref{ADMrigging}) can be equivalently re-expressed in terms of the connection variables, by add the SU(2) Gauss constraint $G_i=\partial_aE^a_i+\epsilon_{ij}^{\ \ k}A_a^jE^a_k$ and the corresponding gauge fixing condition $\xi_i$. The purpose for such a re-expression is to relating the kinematical framework of LQG, which is formulated mathematical-rigorously and better understood than the ADM case
\be
&&\eta_\o: \Fd_{Kin}\to\Fd_{Phys}^\star;\ f\mapsto\eta_\o(f),\ \ \eta_\o(f)[f']:=\nonumber\\
&&\frac{\int \cd E^a_i\cd A_a^i\cd P_I\cd T^I\prod_{x\in M} \delta\Big(G_i\Big)\ \delta\Big(\tilde{C}^{tot}\Big)\ \delta\Big(\tilde{C}^{tot}_j\Big)\ \Delta_{FP}\ \delta\Big(\xi_j\Big)\ e^{ i\int_{t_i}^{t_f}\rmd t\int_{\Sig_t}\rmd^3x\left[E^{a}_i\dot{A}_{a}^i+P_I\dot{T}^I\right]}\ \overline{f\lt(A_a^i,T^I\rt)_{t_f}}f'\lt(A_a^i,T^I\rt)_{t_i}}{\int \cd E^a_i\cd A_a^i\cd P_I\cd T^I\prod_{x\in M} \delta\Big(G_i\Big)\ \delta\Big(\tilde{C}^{tot}\Big)\ \delta\Big(\tilde{C}^{tot}_j\Big)\ \Delta_{FP}\ \delta\Big(\xi_j\Big)\ e^{ i\int_{t_i}^{t_f}\rmd t\int_{\Sig_t}\rmd^3x\left[E^{a}_i\dot{A}_{a}^i+P_I\dot{T}^I\right]}\ \overline{\o\lt(A_a^i,T^I\rt)_{t_f}}\o\lt(A_a^i,T^I\rt)_{t_i}}\label{LQGrigging}
\ee
where $\Delta_{FP}$ is the Faddeev-Popov determinant. In the case that all the kinematical states we considered are SU(2) gauge invariant, we can remove the gauge fixing term $\Delta_{FP}\ \delta\Big(\xi_j\Big)$ from Eq.(\ref{LQGrigging}) without changing anything. The reason is shown by the standard Faddeev-Popov trick, here we briefly outline the method (see also e.g.\cite{Weinberg}):

Suppose a integral formula is written as
\be
Z=\int\cd\mu[X]\ f[X]\ \Delta_{FP}[X]\ \delta\Big(\xi_I[X]\Big)
\ee 
where $\xi_I$ are gauge fixing functions. We replace the variable $X$ everwhere by $X_\L$, which is an finite gauge transformation of $X$ generated by the first-class constraints. Then
\be
Z=\int\cd\mu[X_\L]\ f[X_\L]\ \Delta_{FP}[X_\L]\ \delta\Big(\xi_I[X_\L]\Big)
\ee
We are assuming that the first-class constraint algebra is a Lie algebra so that the gauge transformations form a group. Since the infinitesimal parameter $\L$ is arbitrary, $Z$ cannot dependent on $\L$. Therefore we integral over $\L$ with a certain suitable weight-function $\rho[\L]$
\be
Z\int\cd \L^I\ \rho[\L^I]=\int\cd \L^I\ \rho[\L^I]\int\cd\mu[X_\L]\ f[X_\L]\ \Delta_{FP}[X_\L]\ \delta\Big(\xi_I[X_\L]\Big)
\ee
If we also assume that both the path-integral measure $\cd\mu[X]$ and the function $f[X]$ are invariant under the gauge transformations, we will obtain
\be
Z\int\cd \L^I\ \rho[\L^I]=\int\cd\mu[X]\ f[X] \int\cd \L^I\ \rho[\L^I]\ \Delta_{FP}[X_\L]\ \delta\Big(\xi_I[X_\L]\Big)
\ee
By using an explicit expression of Faddeev-Popov determinant $\Delta_{PF}$. We know that 
\be
\Delta_{PF}[X_\L]=\det\left(\frac{\delta\xi_I[(X_\L)_\l]}{\delta\l^J}\Bigg|_{\l=0}\right)
\ee
We consider the result of performing the gauge transformation with parameters $\L^I$ followed by the gauge transformation with parameters $\l^I$ as a \emph{product} infinitesimal gauge transformation with parameters $\Theta^I(\L,\l)$. Then
\be
\frac{\delta\xi_I[(X_\L)_\l]}{\delta\l^J}\Bigg|_{\l=0}=\frac{\delta\xi_I[X_\Theta]}{\delta\Theta^K}\Bigg|_{\Theta=\L}\frac{\Theta^K(\L,\l)}{\delta\l^J}\Bigg|_{\l=0}=\frac{\delta\xi_I[X_\L]}{\delta\L^K}\ \frac{\Theta^K(\L,\l)}{\delta\l^J}\Bigg|_{\l=0}
\ee
Therefore if we choose the weight-function 
\be
\rho[\L]=\left[\det\left(\frac{\Theta^K(\L,\l)}{\delta\l^J}\Bigg|_{\l=0}\right)\right]^{-1}
\ee
We then have 
\be
\int\cd \L^I\ \rho[\L^I]\ \Delta_{FP}[X_\L]\ \delta\Big(\xi_I[X_\L]\Big)=\int\cd\L^I\ \det\left(\frac{\delta\xi_I[X_\L]}{\delta\L^K}\right)\ \delta\Big(\xi_I[X_\L]\Big)=1
\ee
As a result
\be
Z=\left[\int\cd \L^I\ \rho[\L^I]\right]^{-1}\int\cd\mu[X]\ f[X] 
\ee
which means that $Z$ is expressed as a simpler integral $\int\cd\mu[X]\ f[X]$ divided by a infinite gauge obit volume. Note that if one can define a Haar measure on the gauge group, the weight-function can be chosen as 1. 

The above derivation only depends on two non-trivial assumptions: (1.) Both the measure $\cd\mu$ and the function $f$ are gauge invariant; (2.) The gauge transformations form a Lie group. In our case of Eq.(\ref{LQGrigging}), both integrals in the numerator and denominator fulfill the assumptions. Firstly, in the same way as we discussed in previous sections, the SU(2) gauge transformations are local symmetries of the action and anomaly-free in the path-integral (leave the measure invariant). And we have chosen all the kinematical states $f,f',\o$ are SU(2) gauge invariant. Secondly, the SU(2) gauge transformations form a Lie group. As a result,
\be
&&\eta_\o: \Fd_{Kin}\to\Fd_{Phys}^\star;\ f\mapsto\eta_\o(f),\ \ \eta_\o(f)[f']:=\nonumber\\
&&\frac{\int \cd E^a_i\cd A_a^i\cd P_I\cd T^I\prod_{x\in M} \delta\Big(G_i\Big)\ \delta\Big(\tilde{C}^{tot}\Big)\ \delta\Big(\tilde{C}^{tot}_j\Big)\ e^{ i\int_{t_i}^{t_f}\rmd t\int_{\Sig_t}\rmd^3x\left[E^{a}_i\dot{A}_{a}^i+P_I\dot{T}^I\right]}\ \overline{f\lt(A_a^i,T^I\rt)_{t_f}}f'\lt(A_a^i,T^I\rt)_{t_i}}{\int \cd E^a_i\cd A_a^i\cd P_I\cd T^I\prod_{x\in M} \delta\Big(G_i\Big)\ \delta\Big(\tilde{C}^{tot}\Big)\ \delta\Big(\tilde{C}^{tot}_j\Big)\ e^{ i\int_{t_i}^{t_f}\rmd t\int_{\Sig_t}\rmd^3x\left[E^{a}_i\dot{A}_{a}^i+P_I\dot{T}^I\right]}\ \overline{\o\lt(A_a^i,T^I\rt)_{t_f}}\o\lt(A_a^i,T^I\rt)_{t_i}}\label{LQGrigging1}
\ee
where the gauge fixing terms are removed, and the overall factors of infinite gauge volumes in both numerator and denominator are canceled with each other.  

Another step is to transform the delta functions and express Eq.(\ref{LQGrigging1}) in terms of original constraints $C^{tot}$ and $C^{tot}_a$ 
\be
&&\eta_\o: \Fd_{Kin}\to\Fd_{Phys}^\star;\ f\mapsto\eta_\o(f),\ \ \eta_\o(f)[f']:=\nonumber\\
&&\frac{\int \cd E^a_i\cd A_a^i\cd P_I\cd T^I\prod_{x\in M} R\ \delta\Big(G_i\Big)\ \delta\Big({C}^{tot}\Big)\ \delta\Big({C}^{tot}_a\Big)\ e^{ i\int_{t_i}^{t_f}\rmd t\int_{\Sig_t}\rmd^3x\left[E^{a}_i\dot{A}_{a}^i+P_I\dot{T}^I\right]}\ \overline{f\lt(A_a^i,T^I\rt)_{t_f}}f'\lt(A_a^i,T^I\rt)_{t_i}}{\int \cd E^a_i\cd A_a^i\cd P_I\cd T^I\prod_{x\in M} R\ \delta\Big(G_i\Big)\ \delta\Big({C}^{tot}\Big)\ \delta\Big({C}^{tot}_a\Big)\ e^{ i\int_{t_i}^{t_f}\rmd t\int_{\Sig_t}\rmd^3x\left[E^{a}_i\dot{A}_{a}^i+P_I\dot{T}^I\right]}\ \overline{\o\lt(A_a^i,T^I\rt)_{t_f}}\o\lt(A_a^i,T^I\rt)_{t_i}}\label{LQGrigging2}
\ee
where 
\be
R=\lt|\det\frac{\partial \lt(C^{tot},C^{tot}_a\rt)}{\partial\lt(P_0,P_j\rt)}\rt|=
\frac{1}{\a\sqrt{\det q}}\lt|\det\left(\begin{array}{cc}
{P_0} & {P_j} \\
\partial_aT^0 & \partial_a T^j
\end{array}\right)\rt|
\ee
which depends on both the variables of gravity and the variables of dust. As it was shown generally in section \ref{PIARM}, $\eta_\o(f)$ formally solves the quantum constraint equations defined by the Abelianized constraints Eq.(\ref{csolve}). And 
\be
\lag\eta_\o(f')|\eta_\o(f)\rag_{Phys}:=\eta_\o(f)[f']
\ee
is qualified to be a physical inner product.
 
There is a remarkable feature of Eq.(\ref{LQGrigging2}): when we derive Eq.(\ref{LQGrigging2}), we restrict the kinematical states $f,f',\o$ to be SU(2) gauge invariant. However, even if we apply some gauge non-invariant states to Eq.(\ref{LQGrigging2}), it actually only depends on the equivalence class of the kinematical states $f,f'$, in which different states are related by SU(2) gauge transformations. We can see the reason by a change of variables in the integral, we replace the variables $A_a^i,E^a_i$ by the SU(2) gauge transformed ones $T^\L_GA_a^i,T^\L_GE^a_i$ everywhere (which doesn't change anything), with an arbitrary time-space-dependent parameter $\L(t,\vec{x})$. If we choose the reference vector $\o$ to be gauge invariant, by the fact that $SU(2)$ gauge transformations leaves the measure, delta functions and the kinetic term invariant up to field independent constants, we then obtain that
\be
&&\frac{\int \cd E^a_i\cd A_a^i\cd P_I\cd T^I\prod_{x\in M} R\ \delta\Big(G_i\Big)\ \delta\Big({C}^{tot}\Big)\ \delta\Big({C}^{tot}_a\Big)\ e^{ i\int_{t_i}^{t_f}\rmd t\int_{\Sig_t}\rmd^3x\left[E^{a}_i\dot{A}_{a}^i+P_I\dot{T}^I\right]}\ \overline{f\lt(A_a^i,T^I\rt)_{t_f}}f'\lt(A_a^i,T^I\rt)_{t_i}}{\int \cd E^a_i\cd A_a^i\cd P_I\cd T^I\prod_{x\in M} R\ \delta\Big(G_i\Big)\ \delta\Big({C}^{tot}\Big)\ \delta\Big({C}^{tot}_a\Big)\ e^{ i\int_{t_i}^{t_f}\rmd t\int_{\Sig_t}\rmd^3x\left[E^{a}_i\dot{A}_{a}^i+P_I\dot{T}^I\right]}\ \overline{\o\lt(A_a^i,T^I\rt)_{t_f}}\o\lt(A_a^i,T^I\rt)_{t_i}}\nonumber\\
&=&\frac{\int \cd E^a_i\cd A_a^i\cd P_I\cd T^I\prod_{x\in M} R\ \delta\Big(G_i\Big)\ \delta\Big({C}^{tot}\Big)\ \delta\Big({C}^{tot}_a\Big)\ e^{ i\int_{t_i}^{t_f}\rmd t\int_{\Sig_t}\rmd^3x\left[E^{a}_i\dot{A}_{a}^i+P_I\dot{T}^I\right]}\ \overline{T^{\L_f}_Gf\lt(A_a^i,T^I\rt)_{t_f}}T^{\L_i}_Gf'\lt(A_a^i,T^I\rt)_{t_i}}{\int \cd E^a_i\cd A_a^i\cd P_I\cd T^I\prod_{x\in M} R\ \delta\Big(G_i\Big)\ \delta\Big({C}^{tot}\Big)\ \delta\Big({C}^{tot}_a\Big)\ e^{ i\int_{t_i}^{t_f}\rmd t\int_{\Sig_t}\rmd^3x\left[E^{a}_i\dot{A}_{a}^i+P_I\dot{T}^I\right]}\ \overline{\o\lt(A_a^i,T^I\rt)_{t_f}}\o\lt(A_a^i,T^I\rt)_{t_i}}
\ee
where $\L_i$ and $\L_f$ are relatively independent. This observation shows that for a given kinematical state $f$, $\eta_\o(f)$ solves ALL the quantum constraints, including Gauss constraint, Diffeomorphism constraint and Hamiltonian constraint.

\subsection{The path-integral rigging map in terms of spacetime covariant field variables}

It is more convenient to perform the computation with the path-integral formula in terms of original spacetime covariant field variables and the Lagrangian in a manifestly covariant form. Thus it is better for us to transform Eq.(\ref{LQGrigging2}) into the path-integral in terms of original spacetime covariant field variables, so that we can evaluate the physical inner product by using the techniques of spin-foam model.  

It is shown in the appendix of \cite{EHT} that the canonical path-integral formula from the Ashtekar-Barbero-Immirzi Hamiltonian (the one appeared in Eq.(\ref{LQGrigging2})) is equivalent to the canonical path-integral for the Holst action Eq.(\ref{pi}) after imposing the time-gauge. We incorporate this result to the physical inner product (we first ignore the scalar field contribution, but will add it back afterward):
\be
\lag\eta_\o(f')|\eta_\o(f)\rag_{Phys}&=&\frac{Z_T(f,f')}{Z_T(\o,\o)}\nonumber\\
Z_T(f,f')&=&\int\prod_{x\in M}\left[\rmd A_a^{IJ}(x)\ \rmd\pi_{IJ}^a(x)\ \delta\left(C^{ab}(x)\right)\ \delta\left(D^{ab}(x)\right) \sqrt{\big|\det\Delta_D(x)\big|}\ \right]\ \prod_{x\in M}\left[\delta\left(G^{IJ}(x)\right)\ \delta\Big(H_a(x)\Big)\ \delta\Big(H(x)\Big)\right]\nonumber\\
&&\times\prod_{x\in M}\left[V^{4}_s(x)\ \delta\Big(T_c\Big)\right]\ \exp i\int_{t_i}^{t_f}\rmd t\int_{\Sig_t}\rmd^3x\left[ \frac{1}{2}\pi_{IJ}^a\frac{\partial}{\partial t}\left(A_a^{IJ}-\frac{1}{\gamma}*A_a^{IJ}\right)\right]\ \overline{f\lt(A_a^i\rt)_{t_f}}f'\lt(A_a^i\rt)_{t_i}\label{HP}
\ee 
where the time-gauge condition $T_c:=\eps^{ijk}\pi^a_{0i}\pi^{b}_{jk}\eps_{abc}=0$, $M$ is the spacetime region bounded by initial and final slices $\Sig_{t_i}$ and $\Sig_{t_f}$.

In order to let Eq.(\ref{HP}) consistent with Eq.(\ref{LQGrigging}), one need to be careful for the following reasons: Recall that the simplicity constraint 
\begin{equation}
C^{ab} = \eps^{IJKL}\pi^a_{IJ}\pi^b_{KL} \approx 0
\end{equation}
has five disjoint sectors of solutions, where $\pi^a_{IJ}$ takes one of the five forms
\begin{eqnarray}
(I \pm) && \pi^a_{IJ} = \pm \eps^{abc} e_b^I e_c^J \nonumber\\
(II \pm) && \pi^a_{IJ} = \pm \frac{1}{2} \eps^{abc} e_b^K e_c^L\eps_{IJKL}\nonumber\\
(Deg) && \pi^a_{IJ} = 0.
\end{eqnarray}
If we define $\pi^a_i := \frac{1}{2} \pi^a_{0i}$ and $\tilde{\pi}^a_i:= \frac{1}{4} \epsilon_i{}^{jk} \pi^a_{jk}$. Then we have the following different properties for different sectors
\begin{eqnarray}
(I+) &\Rightarrow& \det \pi^a_i = 0\text{ and }(\det \tilde{\pi}^a_i)(\det e^i_a) > 0, \nonumber\\
(I-) &\Rightarrow& \det \pi^a_i = 0\text{ and }(\det \tilde{\pi}^a_i)(\det e^i_a) < 0, \nonumber\\
(II+) &\Rightarrow& \det \tilde{\pi}^a_i = 0\text{ and }(\det \pi^a_i)(\det e^i_a) > 0, \nonumber\\
(II-) &\Rightarrow& \det \tilde{\pi}^a_i = 0\text{ and }(\det \pi^a_i)(\det e^i_a) < 0 .
\end{eqnarray}
where we can also see that the four sectors $(I \pm)$ and $(II \pm)$ are disjoint. Because of the appearance of these five sectors of solutions, we have to clarify which sector is contained in the integral of Eq.(\ref{HP}) in order to be consistent with Eq.(\ref{LQGrigging}). It turns out in the appendix of \cite{EHT} that the integral of Eq.(\ref{HP}) only contain two sectors, either $(I\pm)$ or $(II\pm)$. Thus we restrict ourselves in the sectors $(II\pm)$.

The next step is to perform the Henneaux-Slavnov trick \cite{BHNR} to eliminate the secondary second-class constraint $D^{ab}$ in the path-integral formula. We express the delta functions $\delta(H)$ and $\delta(D^{ab})$ by their Fourier decompositions,
\be
Z_T(f,f')&=&\int_{II\pm} \prod_{x\in M}\left[\rmd A_a^{IJ}(x)\ \rmd\pi_{IJ}^a(x)\ \rmd d^{ab}(x)\ \delta\left(C^{ab}(x)\right)\ \sqrt{\big|\det\Delta_D(x)\big|}\ \right]\ \prod_{x\in M}\left[\rmd N(x)\ \delta\left(G^{IJ}(x)\right)\ \delta\Big(H_a(x)\Big)\right]\nonumber\\
&&\times\prod_{x\in M}\left[V^{4}_s(x)\ \delta\Big(T_c\Big)\right]\ \exp i\int_{t_i}^{t_f}\rmd t\int_{\Sig_t}\rmd^3x\left[ \frac{1}{2}\pi_{IJ}^a\frac{\partial}{\partial t}\left(A_a^{IJ}-\frac{1}{\gamma}*A_a^{IJ}\right)-NH+d_{ab}D^{ab}\right]\ \overline{f\lt(A_a^i\rt)_{t_f}}f'\lt(A_a^i\rt)_{t_i}
\ee
Then We consider a change of variables which is also a canonical transformation for the canonical fields on different spatial slices $\Sig_t$. It is generated by the functional 
\be
F(t)&:=&-\int_{\Sig_t}\rmd^3x\ \tilde{d}_{ab}(t,\vec{x})\ C^{ab}/N(t,\vec{x})\nonumber\\
\text{where}\ \ \ \ \ \ \ \ \ \ \ \ \ 
\tilde{d}_{ab}(t,\vec{x}) 
&=&{d}_{ab}(t,\vec{x})\ \ \ \ \ \ \ \text{for}\ t\in[t_i+\eps,t_f-\eps]\nonumber\\
&=&0\ \ \ \ \ \ \ \ \ \ \ \ \ \ \ \ \ \text{for}\ t=\{t_i,t_f\}
\ee
so this change of variable doesn't affect the kinematical state $f,f'$ on the boundary. The integral measure $\left[\rmd A_a^{IJ}(x)\ \rmd\pi_{IJ}^a(x)\right]$ is the Liouville measure on the phase space thus is invariant under this canonical transformation, $\sqrt{|\det\Delta_D(x)|}$ is also invariant under the canonical transformation generated by $F$ since $\left\{C^{ab}(x), G^{cd,ef}(x',x'')\right\}=0$, and the product of $\delta$-functions for both time-gauge and first-class constraints are invariant under this canonical transformation because 
\be
\left\{G_{IJ}(\L^{IJ}), C^{ab}(c_{ab})\right\}&=&0\nonumber\\
\left\{C^{ab}(c_{ab}),C^{cd}(d_{cd})\right\}&=&0\nonumber\\
\left\{H_a(N^a),C^{bc}(c_{bc})\right\}&=&C^{ab}(\cl_{\vec{N}}c_{ab})
\ee
Under the canonical transformation generated by $F$, the change of kinetic term $\delta\int\rmd t\int\rmd^3x\ \pi_{IJ}^a\partial_t\left(A_a^{IJ}-\frac{1}{\gamma}*A_a^{IJ}\right)$ is proportional to $\int\rmd t\int\rmd^3x\ C^{ab}\partial_t(d_{ab}/N)$ which also vanishes by the delta functions $\delta(C^{ab})$ in front of the exponential. So $H$ and $D^{ab}$ are the only terms variant in this canonical transformation. Moreover because $\left\{H(x),C^{ab}(x')\right\}=D^{ab}(x)\delta(x,x')$ modulo the terms proportional to $C^{ab}$ and $\left\{C^{ab}(x),D^{cd}(x')\right\}=G^{ab,cd}(x,x')$ we can obtain explicitly the transformation behavior of $H(N)$ and $D^{cd}(d_{cd})$ in the time period $[t_i+\eps,t_f-\eps]$ modulo the terms vanishing on the constraint surface defined by $C^{ab}=0$
\be
e^{\hat{\chi}_{-F}}{H}(N)&\equiv&\sum_{n=0}^\infty\frac{1}{n!}\{F,H(N)\}_{(n)}
\ =\ \int\rmd^3x\ N(x)H(x)+\int\rmd^3x\ d_{ab}(x)D^{ab}(x)-\frac{1}{2}\int\rmd^3y\int\rmd^3z\frac{1}{N(y)}d_{ab}(y)d_{cd}(z)G^{ab,cd}(y,z)\nonumber\\
e^{\hat{\chi}_{-F}}{D}^{cd}(d_{cd})&\equiv&\sum_{n=0}^\infty\frac{1}{n!}\{F,D^{cd}(d_{cd})\}_{(n)}
\ =\ \int\rmd^3x\ d_{cd}(x)D^{cd}(x)-\int\rmd^3x\int\rmd^3y\ \frac{1}{N(y)}d_{cd}(x)d_{ab}(y)G^{ab,cd}(x,y)
\ee
here $\chi_{-F}$ is the Hamiltonian vector field associated by the phase space function $-F$, and the series terminated because of 
\be
\left\{C^{ab}(x), G^{cd,ef}(x',x'')\right\}=0.
\ee 
When we take $\eps\to0$, the integral on the exponential becomes 
\be
\int\rmd t\int\rmd^3x\left[ \frac{1}{2}\pi_{IJ}^a\frac{\partial}{\partial t}\left(A_a^{IJ}-\frac{1}{\gamma}*A_a^{IJ}\right)-NH-\frac{1}{2}d_{ab}d_{cd}G^{ab,cd}/N\right]
\ee
Then we perform the integral over $d_{ab}$, we obtain
\be
&&Z_T(f,f')\nonumber\\
&=&\int_{II\pm} \left[\cd A_a^{IJ} \cd \pi_{IJ}^a\ \delta\left(C^{ab}\right) \sqrt{|\det G|}\ \right]\ \left[\cd A_t^{IJ} \cd N^a \cd N\ \lt|N^3\rt|\ \right]\prod_{x\in M}\left[V^{4}_s\ \delta\Big(T_c\Big)\right] \nonumber\\
&&\times \exp{i\int_{t_i}^{t_f}\rmd t\int_{\Sig_t}\rmd^3x\left[ \frac{1}{2}\pi_{IJ}^a{\partial_t}\left(A_a^{IJ}-\frac{1}{\gamma}*A_a^{IJ}\right)-\frac{1}{2}\left(A_t^{IJ}+N^aA_a^{IJ}\right)G_{IJ}-N^aH_a-NH\right]}\ \overline{f\lt(A_a^i\rt)_{t_f}}f'\lt(A_a^i\rt)_{t_i}\label{HP1}
\ee
We include the scalar degrees of freedom, since the scalar field contributions all commute with $F(t)$, we get
\be
&&Z_T(f,f')\nonumber\\
&=&\int_{II\pm} \left[\cd A_a^{IJ} \cd \pi_{IJ}^a\ \delta\left(C^{ab}\right) \sqrt{|\det G|}\ \right]\ \left[\cd A_t^{IJ} \cd N^a \cd N\ \lt|N^3\rt|\ \right]\lt[\cd P_I\cd T^I\rt]\prod_{x\in M}\left[V^{4}_s\ \delta\Big(T_c\Big)\right]\ R\lt[P_I,T^I,q^{ab}\rt] \nonumber\\
&&\times \exp i\int_{t_i}^{t_f}\rmd t\int_{\Sig_t}\rmd^3x\Bigg[ \frac{1}{2}\pi_{IJ}^a{\partial_t}\left(A_a^{IJ}-\frac{1}{\gamma}*A_a^{IJ}\right)-\frac{1}{2}\left(A_t^{IJ}+N^aA_a^{IJ}\right)G_{IJ}-N^aH_a-NH\nonumber\\
&&+P_I{\partial_t}T^I-N^aP_I\partial_aT^I-\frac{N}{\sqrt{\det q}}\lt(\frac{\a(\det q)}{2}q^{ab}\partial_aT^I\partial_bT^I+\frac{1}{2\a}P_I^2\rt)\Bigg]\nonumber\\
&&\times\overline{f\lt(A_a^i,T^I\rt)_{t_f}}f'\lt(A_a^i,T^I\rt)_{t_i}\label{HP2}
\ee
We then consider the scalar field contributions of the total action on the exponential:
\be
S_{KG}&=&\int_{t_i}^{t_f}\rmd t\int_{\Sig_t}\rmd^3x\Bigg[P_I{\partial_t}T^I-N^aP_I\partial_aT^I-\frac{N}{\sqrt{\det q}}\lt(\frac{\a(\det q)}{2}q^{ab}\partial_aT^I\partial_bT^I+\frac{1}{2\a}P_I^2\rt)\Bigg]\nonumber\\
&=&\int_{t_i}^{t_f}\rmd t\int_{\Sig_t}\rmd^3x\Bigg[-\frac{N}{2\a\sqrt{\det q}}P_I^2+NP_In^\a{\partial_\a}T^I-\frac{\a}{2}{N}{\sqrt{\det q}}q^{ab}\partial_aT^I\partial_bT^I\Bigg]\nonumber\\
&=&\int_{t_i}^{t_f}\rmd t\int_{\Sig_t}\rmd^3x\Bigg[-\frac{N}{2\a\sqrt{\det q}}P_I^2+NP_In^\a{\partial_\a}T^I-\frac{\a}{2}N{\sqrt{\det q}}\lt( n^\a{\partial_\a}T^I\rt)^2+\frac{\a}{2}N{\sqrt{\det q}}\lt( n^\a{\partial_\a}T^I\rt)^2\nonumber\\
&&-\frac{\a}{2}{N}{\sqrt{\det q}}q^{ab}\partial_aT^I\partial_bT^I\Bigg]\nonumber\\
&=&-\int_{t_i}^{t_f}\rmd t\int_{\Sig_t}\rmd^3x\Bigg[\sqrt{\frac{N}{2\a\sqrt{\det q}}}P_I-\sqrt{\frac{\a}{2}N{\sqrt{\det q}}}\lt( n^\a{\partial_\a}T^I\rt)\Bigg]^2-\frac{\a}{2}\int\rmd^4x\ \sqrt{|\det(g)|}\ g^{\a\b}\partial_\a T^I \partial_\b T^I
\ee
where the second term is the original covariant action of the scalar fields, and the first term contributes the integral of $P_I$. Let's extract the integral of $P_I$ out of the path-itnegral formula:
\be
&&\int\cd P_I\ R\ e^{-i\int_{t_i}^{t_f}\rmd t\int_{\Sig_t}\rmd^3x\lt[\sqrt{\frac{N}{2\a\sqrt{\det q}}}P_I-\sqrt{\frac{\a}{2}N{\sqrt{\det q}}}\lt( n^\a{\partial_\a}T^I\rt)\rt]^2}\nonumber\\
&=&\frac{1}{\a\sqrt{\det q}}\int\cd P_I\ \lt|\det\left(\begin{array}{cc}
{P_0} & {P_j} \\
\partial_aT^0 & \partial_a T^j
\end{array}\right)\rt|\ e^{-i\int_{t_i}^{t_f}\rmd t\int_{\Sig_t}\rmd^3x\lt[\sqrt{\frac{N}{2\a\sqrt{\det q}}}P_I-\sqrt{\frac{\a}{2}N{\sqrt{\det q}}}\lt( n^\a{\partial_\a}T^I\rt)\rt]^2}\nonumber\\
&=&\frac{1}{\a\sqrt{\det q}}\lt(\frac{2\a\sqrt{\det q}}{N}\rt)^{5/2}\int\cd X_I\ \lt|\det\left(\begin{array}{cc}
{X_0+\sqrt{\frac{\a}{2}N{\sqrt{\det q}}}\ n^\a{\partial_\a}T^0} & {X_j+\sqrt{\frac{\a}{2}N{\sqrt{\det q}}}\ n^\a{\partial_\a}T^j} \\
\partial_aT^0 & \partial_a T^j
\end{array}\right)\rt|\ e^{-i\int_{t_i}^{t_f}\rmd t\int_{\Sig_t}\rmd^3x\ X_I^2}\nonumber\\
&\equiv&2^{5/2}\a^{3/2}V_s^{3/2}N^{-5/2}\cj_{KG}\lt[\a,T^I,q^{ab}\rt]\label{JKG}
\ee

After the above manipulation for the scalar field degrees of freedom, we follow the same way in \cite{EHT} to solve the simplicity constraint $C^{ab}$, and obtain a path-integral of the Holst action coupling to the scalar fields
\be
Z_T(f,f')&=&\int_{II\pm} \cd A_\a^{IJ} \cd e^I_\a\cd T^I \prod_{x\in M}\lt|\cv^{1/2} V_s^6\rt|\ \cj_{KG}\lt[\a,T^I,q^{ab}\rt]\ \delta^3\lt(e^0_a\rt)\ \lt[\sin\int_M\ e^I\wedge e^J\wedge\left(* F_{IJ}-\frac{1}{\gamma}F_{IJ}\right)\rt]\nonumber\\ 
&&\times \exp\lt[-\frac{i\a}{2}\int_M\rmd^4x\ \sqrt{|\det(g)|}\ g^{\a\b}\partial_\a T^I \partial_\b T^I\rt]\overline{f\lt(A_a^i,T^I\rt)_{t_f}}f'\lt(A_a^i,T^I\rt)_{t_i}\label{HSSM}
\ee
It is the path-integral representation of the physical inner product for GR coupling to the scalar fields. However, the Jacobian $\cj_{KG}$ involves integral expression Eq.(\ref{JKG}) thus is hard to practically compute. But we can approximate this formula Eq.(\ref{HSSM}) by the pure gravity contribution, if we assume 
(1.) $f,f',\o$ depend on $A_a^i$ only; 
(2.) the coupling constant $\a$ is small and negligible. 
The reason is that if $\a$ is negligible, $\cj_{KG}$ then becomes independent of $q^{ab}$, then the integrals of scalar field variables are factored out from Eq.(\ref{HSSM}) and canceled between the denominator and the numerator of the physical inner product:
\be
\lag\eta_\o(f')|\eta_\o(f)\rag_{Phys}&=&\frac{Z_T(f,f')}{Z_T(\o,\o)}
\ee
Under this approximation, we write down the path-integral representation of the physical inner product
\be
Z_T(f,f')=\int_{II\pm} \cd A_\a^{IJ} \cd e^I_\a \prod_{x\in M}\lt|\cv^{1/2} V_s^6\rt|\ \delta^3\lt(e^0_a\rt)\ \lt[\sin\int_M\ e^I\wedge e^J\wedge\left(* F_{IJ}-\frac{1}{\gamma}F_{IJ}\right)\rt]\ \overline{f\lt(A_a^i\rt)_{t_f}}f'\lt(A_a^i\rt)_{t_i}\label{HSM}
\ee
Moreover, Eq.(\ref{HSM}) is equivalent to the path-integral of a Plebanski-Holst action \cite{EHT}
\be
Z_T(f,f')&=&\int_{II\pm}\cd A_\a^{IJ}\cd B_{\a\b}^{IJ} \prod_{x\in M} \lt|\cv^{13/2} V^{9}_s\rt|\ \delta^{20}\left(\eps_{IJKL}\ B_{\a\b}^{IJ}\ B_{\g\delta}^{KL}-\frac{1}{4!}\cv\eps_{\a\b\g\delta}\right)\ \delta^3(T_c)\nonumber\\
&&\times\lt[\exp i\int_M B^{IJ}\wedge (F-\frac{1}{\g}* F)\rt]\ \overline{f\lt(A_a^i\rt)_{t_f}}f'\lt(A_a^i\rt)_{t_i}.\label{PSM}
\ee

\section{Conclusion and discussion}

The aim of the present paper has been to analyze the gauge invariance of the path-integral measure consistent with the canonical settings, also to give a path-integral formula appropriate for both conceptual and practical purposes. Our previous discussions show that the path-integral measure we obtained is invariant under all the gauge transformations (local symmetries) of GR generated by the first-class constraints. These gauge transformations form the Bergmann-Komar ``group" (enveloping algebra) which is the collection of the dynamical symmetries. We also obtain the desired path-integral formula which formally solves all the constraints of GR quantum mechanically. And the pure gravity part of physical inner product is formally represented by these path-integral formula Eqs.(\ref{HSM}) or (\ref{PSM}), which is ready for the spin-foam model construction.
 
There are several remarks concerning our result Eqs.(\ref{HSM}) and (\ref{PSM}):
\begin{itemize}
\item In Eqs.(\ref{HSM}) and (\ref{PSM}) most of the gauge fixing conditions, which often appear in the quantization of gauge system, disappear in our case. Since it is hard to implement gauge fixing conditions in the spin-foam quantization, the Eqs.(\ref{HSM}) and (\ref{PSM}) simplify the further construction and computation at this point. However, the appearance of time-gauge will have non-trivial contribution to the construction of a spin-foam model.

\item Our construction formally explains that if we are working toward formally a rigging map and physical inner product consistent with the Ashtekar-Barbero-Immirzi canonical formulation \cite{ABI}, we have to include both two sectors $(II\pm)$ in constructing the spin-foam models from Plebanski path-integral. It simplifies the construction because in constructing a spin-foam model, we would like to first remove the delta functions of the simplicity constraint
$
\eps_{IJKL}\ B_{\a\b}^{IJ}\ B_{\g\delta}^{KL}=\frac{1}{4!}\cv\eps_{\a\b\g\delta}
$
and consider a pure BF-theory, and implement the simplicity constraint afterwards. In considering BF theory, we need the integral for each component of $B_{\a\b}^{IJ}$ to be over the full real-line in order to obtain a product of $\delta\lt(F\rt)$ (more precisely the delta functions of holonomies). However if it was restricted to only one single sector $(II+)$ or $(II-)$, the integral for each component of $B_{\a\b}^{IJ}$ would be only over a half real-line.

\item The local measure factors ($\cv^{1/2} V_s^6$ for Holst and $\cv^{13/2} V^{9}_s$ for Plebanski-Holst) appear in our result Eqs.(\ref{HSM}) and (\ref{PSM}). It means that in order to interpret the path-integral amplitude (or spin-foam amplitude) as a physical inner product in the canonical theory, one has to properly implement these local measure factors into the spin-foam model. The detailed implementation and construction will appear in the future publication. 

\item In order to construct the physical Hilbert space $\ch_{Phys}:=\eta_{\o}(\Fd_{Kin})/\Fn$, it is needed to clarify the null space 
\be
\Fn=\lt\{\eta_\o(f)\in\Fd_{Phys}^\star\ \big|\ ||\eta_\o(f)||_{Phys}=0\ \rt\}
\ee 
implied by the physical inner product defined by Eqs.(\ref{HSM}) or (\ref{PSM}). Therefore the further research is necessary regarding the implication from the path-integral formula Eqs.(\ref{HSM}) or (\ref{PSM}).

\end{itemize}

\section*{Acknowledgments}

M.H. is grateful for the advises from Thomas Thiemann and the fruitful discussion with Aristide Baratin, Bianca Dittrich, and Jonathan Engle. M.H. also would like to gratefully acknowledge the support by International Max Planck Research School and the partial support by NSFC Nos. 10675019 and 10975017.

\end{document}